\documentclass[sigconf]{acmart}

\AtBeginDocument{%
  \providecommand\BibTeX{{%
    \normalfont B\kern-0.5em{\scshape i\kern-0.25em b}\kern-0.8em\TeX}}}

\setcopyright{acmlicensed}
\copyrightyear{2024}
\acmYear{2024}
\acmDOI{2404.02718}

\acmConference[xxxx]{xxxx}{xxxxx}{xxx, xxxx}
\begin{document}

%%
%% The "title" command has an optional parameter,
%% allowing the author to define a "short title" to be used in page headers.
\title{Evolving Agents: Interactive Simulation of Dynamic and Diverse Human Personalities}

%%
%% The "author" command and its associated commands are used to define
%% the authors and their affiliations.
%% Of note is the shared affiliation of the first two authors, and the
%% "authornote" and "authornotemark" commands
%% used to denote shared contribution to the research.
\author{Jiale Li}
\authornote{Both authors contributed equally to this research.}
\email{2233676@tongji.edu.com}
\affiliation{%
  \institution{Tongji University}
  \city{Shanghai}
  \country{China}
}

\author{Jiayang Li}
\authornotemark[1]
\email{2233666@tongji.edu.cn}
\affiliation{%
  \institution{Tongji University}
  \city{Shanghai}
  \country{China}
}

 \author{Jiahao Chen}
 \affiliation{%
   \institution{Tongji University}
   \city{Shanghai}
   \country{China}
 }

 \author{Yifan Li}
 \affiliation{%
   \institution{Tongji University}
   \city{Shanghai}
   \country{China}
 }

 \author{Shijie Wang}
 \affiliation{%
   \institution{Tongji University}
   \city{Shanghai}
   \country{China}
 }

 \author{Hugo Zhou}
 \affiliation{%
   \institution{Tongji University}
   \city{Shanghai}
   \country{China}
 }

 \author{Minjun Ye}
 \affiliation{%
   \institution{Tongji University}
   \city{Shanghai}
   \country{China}
 }

 \author{Yunsheng Su}
 \authornote{Corresponding author}
 \affiliation{%
   \institution{Tongji University}
   \city{Shanghai}
   \country{China}
 }

%%
%% By default, the full list of authors will be used in the page
%% headers. Often, this list is too long, and will overlap
%% other information printed in the page headers. This command allows
%% the author to define a more concise list
%% of authors' names for this purpose.

% \renewcommand{\shortauthors}{Jiale Li and Jiayang Li, et al.}
%%
%% The abstract is a short summary of the work to be presented in the
%% article.

%%
%The code below is generated by the tool at http://dl.acm.org/ccs.cfm.
%% Please copy and paste the code instead of the example below.
%%
\begin{abstract}
    
    Human-like Agents with diverse and dynamic personalities could serve as an important design probe in the user-centred design process, thereby enabling designers to enhance the user experience of interactive applications.

    In this article, we introduce Evolving Agents, a novel agent architecture that consists of two systems: Personality and Behavior. The Personality system includes Cognition, Emotion, and Character Growth modules. The Behavior system comprises two modules: Planning and Action. We also build a simulation platform that enables agents to interact with the environment and other agents. 

    Evolving Agents can simulate the human personality evolution process. Compared to its initial state, agents' personalities and behavior patterns undergo believable development after several days of simulation. Agents reflect on their behavior to reason and develop new personality traits. These traits, in turn, generate new behavior patterns, forming a feedback loop-like personality evolution.

    Our experiment utilized a simulation platform with ten agents for evaluation. During the evaluation, these agents experienced believable and inspirational personality evolution. We demonstrated the effectiveness of agent personality evolution through ablation and control experiments. All of our agent architecture modules contribute to creating believable human-like agents with diverse and dynamic personalities. We also demonstrated through workshops how Evolving Agents could inspire designers.

\end{abstract}

\begin{CCSXML}
<ccs2012>
 <concept>
  <concept_id>00000000.0000000.0000000</concept_id>
  <concept_desc>Human-centered computing -> Interactive systems and tools;·Computing methodologies -> Natural language prcessing</concept_desc>
  <concept_significance>500</concept_significance>
 </concept>
 <concept>
  <concept_id>00000000.00000000.00000000</concept_id>
  <concept_desc>Do Not Use This Code, Generate the Correct Terms for Your Paper</concept_desc>
  <concept_significance>300</concept_significance>
 </concept>
 <concept>
  <concept_id>00000000.00000000.00000000</concept_id>
  <concept_desc>Do Not Use This Code, Generate the Correct Terms for Your Paper</concept_desc>
  <concept_significance>100</concept_significance>
 </concept>
 <concept>
  <concept_id>00000000.00000000.00000000</concept_id>
  <concept_desc>Do Not Use This Code, Generate the Correct Terms for Your Paper</concept_desc>
  <concept_significance>100</concept_significance>
 </concept>
</ccs2012>
\end{CCSXML}

\ccsdesc[500]{Human-centered computing~Interactive systems and tools}
\ccsdesc[500]{Computing methodologies~Natural language processing}

%%
%% Keywords. The author(s) should pick words that accurately describe
%% the work being presented. Separate the keywords with commas.
\keywords{Human-AI interaction, Human-Like agents, generative AI, large language models, cognitive psychology, personality psychology}

%% A "teaser" image appears between the author and affiliation
%% information and the body of the document, and typically spans the
%% page.
\begin{teaserfigure}
  \includegraphics[width=\textwidth]{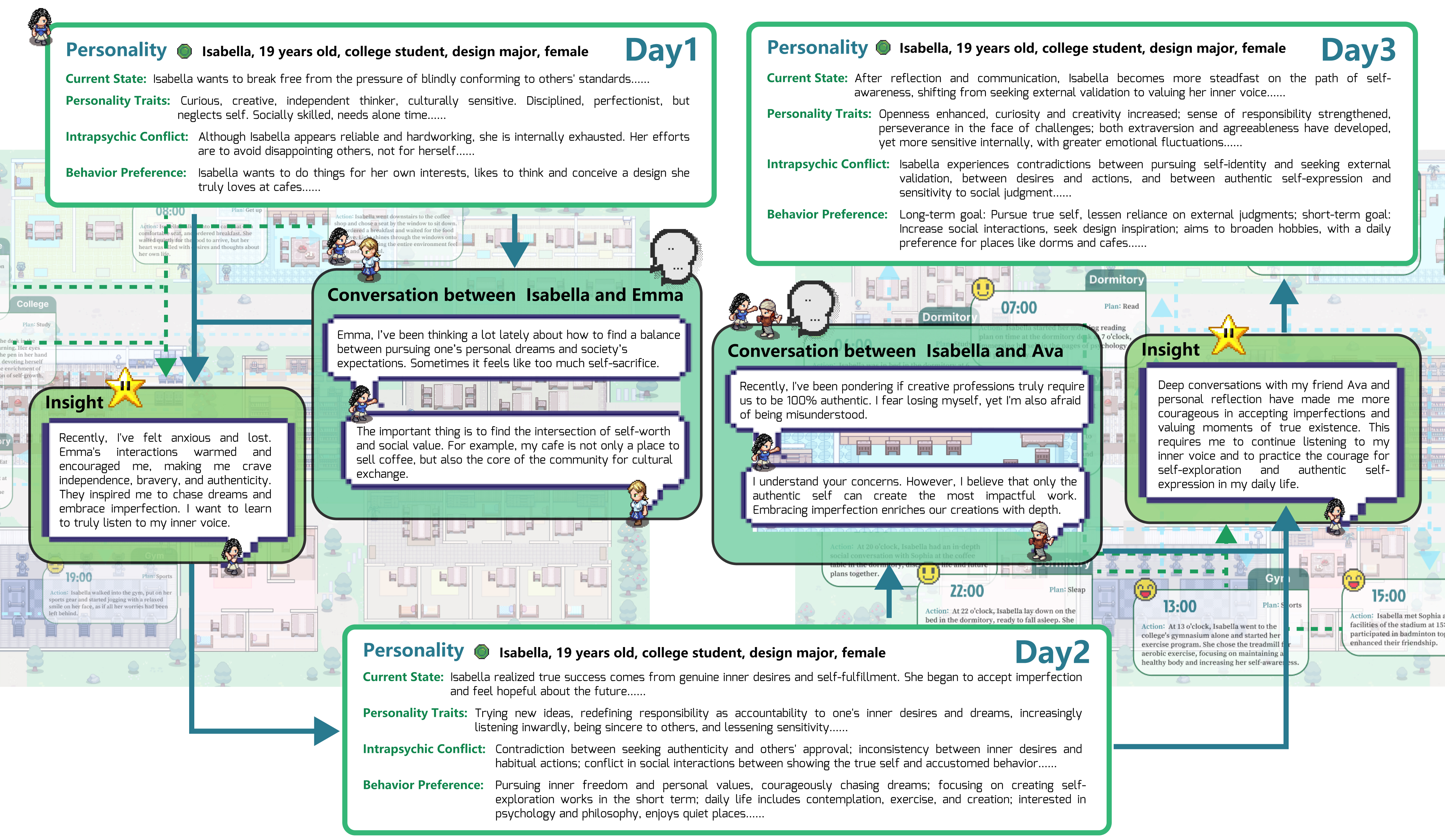}
  \caption{Evolving Agents could exhibit differentiated behaviors corresponding to their personalities and 
undergo continuous personality evolution based on external information during the interactive simulation system}
  \Description{Enjoying the baseball game from the third-base
  seats. Ichiro Suzuki is preparing to bat.}
  \label{fig:teaser}
\end{teaserfigure}

% \received{20 February 2007}
% \received[revised]{12 March 2009}
% \received[accepted]{5 June 2009}

%%
%% This command processes the author and affiliation and title
%% information and builds the first part of the formatted document.
\maketitle

\section{INTRODUCTION}
Embodied agents are capable of simulating the human's behaviors. By observing and interacting with these human-like agents, we can better understand ourselves as humans. Within a design context, analyzing how these agents operate in specific scenarios allows designers to uncover insights into potential user behaviors, needs, and experiences\cite{hong_conceptualization_2017,mcginn_datadriven_2008}, making the embodied agents an inspirational design probe in the user-centred design process.

Numerous studies have proposed that embodied agents can simulate human-like behaviors within virtual environments. Agents can actively interact with objects and other agents in the simulation environment\cite{noauthor_generative_nodate,wang_humanoid_2023,lin_agentsims_2023} or continuously learn new behaviors guided by predefined tasks in the environment\cite{wang_voyager_2023,fan_minedojo_2022}.

One of the foremost objectives in the development of embodied agents is to enhance their behavior's similarity to human actions within the simulation environment\cite{zhu_ghost_2023,chen_llm-empowered_2023,du_improving_2023,zhang_building_2024}. Real humans exhibit a wide array of personality traits and corresponding behavior patterns\cite{zuckerman_psychobiology_1991} that dynamically evolve through interacting with the external world\cite{ghasem-aghaee_cognitive_2007}. However, current embodied agents show a constrained range of personality expressions. However, their behaviors and personalities do not undergo evolution over time via continuous interaction with the simulation environments. In previous sandbox simulation studies, there are seldom instances of the personality evolution process occurring without task-oriented guidance. Agents that do not undergo continuous evolution seem to experience each day as one long moment, repeating itself repeatedly.

Furthering this point, current agents rarely exhibit diversified and dynamic personalities within the interactive simulation process. This results in a lack of variation in behavioral patterns both among different agents and across different simulation periods of the same Agent. Therefore, we aim to design a novel agent architecture that enables agents to possess their own diversified and dynamic personalities, which could make agents experience their days as thousands of single and different moments, which is called personality evolution in previous work\cite{oren_dynamic_1991}.

To realize agents' personality evolution, enabling agents to exhibit differentiated behaviors corresponding to their various personalities and undergo continuous personality evolution based on external information, we drew inspiration from psychological theories\cite{lachman_cognitive_2016,kihlstrom_cognitive_2018,allport_pattern_1961,baars_cognitive_1993}. We proposed a novel architecture - Evolving Agents. Evolving Agents can simulate believable personality traits and personalized behavioral patterns through simulation and experimental analysis in sandbox environments. Moreover, their personalities and behaviors undergo continuous evolution throughout the simulation process. Evolving Agents exhibit differentiated behavioral patterns based on their diverse personalities. At the same time, their behavioral experiences also influence the changes in agent personalities simultaneously, forming a feedback loop-like process of behavior-personality development\cite{brun_engineering_2009}. For example, Benjamin is committed to changing his Introverted personality and tries to engage in more interpersonal communication. His enjoyable social experiences further enhance Benjamin's extroversion, leading to a more open-minded attitude toward interpersonal interactions. Similarly, Isabella actively explores her values during the process, and the encouragement she receives from others reinforces her determination to live for herself, consequently enhancing Isabella's self-awareness.

Evolving Agents' architectural design incorporates insights from psychology and computer science. By adopting the stimulus-response concept from behavioral psychology\cite{clark_learning_2018,woodward_discovery_1982}, we have developed two main systems: the Behavior System, responsible for planning specific behaviors and execution. The Personality system, constructed with Emotion, Cognition, and Character Growth modules, cognitively processes external information to enable continuous evolution. The interaction between these two systems allows agents to exhibit human-like personalized behavior patterns and undergo dynamic personality evolution. Additionally, we have developed a new sandbox environment where agents can interact with environmental scenarios and other agents. This platform allows users to edit environments and interact with agents, providing intuitive and engaging simulations.

To validate the perception and believability of Evolving Agents simulating diverse human personality traits and dynamic personality evolution, we proposed several research hypotheses:
\begin{itemize}
    \item Evolving Agents can construct agents with perceptible and believable differentiated personality traits.
    \item Evolving Agents' personality descriptions undergo perceptible and believable evolution during interactive simulations.
    \item Evolving Agents' dynamic personality evolution can induce perceptible and believable changes in their behavior.
\end{itemize}

We conducted objective data analysis and subjective human evaluation, with both results preliminarily confirming our research hypotheses. In the objective data analysis, we utilized the Big Five personality traits scale and behavioral similarity analysis to verify whether Evolving Agents have undergone differentiated and dynamic personality evolution and whether the behavior pattern develops continuously. In human evaluation, we engaged 30 human assessors to evaluate the believability of personality differences and the personality evolution in Evolving Agents. They were also asked to compare several ablations restricting the agents' access to the Character Growth module and complete Character Structure. These demonstrate that Evolving Agents can perceptibly and believably exhibit differentiated behaviors in alignment with their personalities and undergo continuous and believable personality evolution. Each of these components is critical to strong performance. Furthermore, we organized designer workshops around this system to discuss the inspiring effect of Evolving Agents as design probes for designers, yielding valuable insights into integrating such systems into user-oriented design processes.

In sum, this paper makes the following contributions: 
\begin{itemize}
    \item Evolving Agents, perceptible and believable simulacra of diverse and dynamic human personalities in the interactive simulation system.
    \item A novel agent architecture, having behavior and personality as two main systems, makes it possible to execute planning, Action, emotion, cognition, and character growth in the Virtual Simulation Environment, enabling agents to exhibit differentiated behaviors corresponding to their various personalities and undergo continuous personality evolution based on external information during the interactive simulation.
    \item An evaluation procedure was established that includes two methods, objective data verification, and human evaluation, for assessing the effectiveness of agent personality evolution and the contributions of each module within the architecture during this process. Additionally, we conducted a supportive workshop to discuss the inspirational effects of Evolving Agents on designers.
\end{itemize}
\section{RELATED WORK}
\subsection{Interaction Between Humans And Agents}
As generative models advance rapidly in general capabilities and intent recognition\cite{zuckerman_psychobiology_1991,hamilton_blind_2023,swan_math_2023,wei_emergent_2022,ouyang_training_2022}, agents powered by these models have found widespread application across various fields and task types\cite{xi_rise_2023,wei_emergent_2022,sumers_cognitive_2024,liu_training_2023}, demonstrating a broad spectrum of application prospects. With the enrichment of application scenarios for agents, novel patterns of interaction between humans and agents are continually emerging\cite{kim_one_2023,li_map_2024,wang_human-human_2020}.

Human-agent interaction, as the name suggests, involves agents collaborating with humans to accomplish tasks\cite{xi_rise_2023}, such as introducing agents to assist in peer psychological counseling\cite{hsu_helping_2023}, using agents in telemarketing scenarios\cite{noauthor_multi-turn_nodate}, or co-creating literature or artworks with agents\cite{bhat_interacting_2023,noauthor_3dall-e_nodate,chiou_designing_2023}.

In recent years, the human-like behavior exhibited by embodied agents in simulation environments has attracted widespread social attention\cite{kalvakurthi_hey_2023,hsu_helping_2023,hasan_sapien_2023,noauthor_human-level_nodate,schick_peer_2023}. Among these, the Generative Agent discussed human social and communication behaviors within societies. By observing simulated behaviors of agents in a sandbox environment\cite{noauthor_generative_nodate}, we can know ourselves as humans better.

Some research focuses on multi-agent simulation in more goal-oriented settings, such as collaborative efforts in software production within software studios\cite{qian_communicative_2023-1}. At the same time, some others focused on creating agents capable of autonomous exploration and continuous learning within game worlds\cite{wang_voyager_2023}. Additionally, studies aim to develop human-like NPCs, enabling text-based strategy game interactions with users and providing imaginative possibilities for future interaction between game players and NPCs or companions.

When focusing on the interaction between designers and agents, designers can observe the simulated behaviors and feedback of agents representing target users, gaining valuable design analysis materials to assist in the user-oriented design process\cite{zamfirescu-pereira_why_2023,zamfirescu-pereira_herding_2023}. Current research has explored this direction; for example, one study aimed to simulate doctor-patient communication scenarios in psychological counselling to provide references and suggestions for real treatment\cite{chen_llm-empowered_2023}. Another study simulated user behaviors in scenarios like online movie ratings, offering insights for further exploration in these areas\cite{hong_conceptualization_2017}.
\subsection{Human-like Agents In Virtual Simulation Environment}
Simulation of human behavior in the real-world environment has been an interesting and important area for a long time\cite{noauthor_computer_nodate,cote_textworld_2019,hausknecht_interactive_2020,wang_describe_2023,nottingham_embodied_2023}. Ushered in by the landmark paper on Generative Agents\cite{park_generative_2023}, the promise of modelling conceivable human-like behavior in Virtual Simulation Environment has sparked the imagination of many\cite{noauthor_230812503_nodate,li_hitchhikers_2023,wang_scienceworld_2022}.

As informed by sociological opinions, the human-like dimension of embodied agents can be subdivided into intrinsic and extrinsic aspects\cite{noauthor_becoming_nodate,xi_rise_2023}. The extrinsic aspect encompasses behaviors at both the individual and group levels\cite{liu_llmp_2023,qian_communicative_2023-1,chan_chateval_2023,chen_agentverse_2023}. In contrast, the intrinsic aspect comprises emotions, characteristics, and cognition\cite{tu_characterchat_2023,jiang_evaluating_2023,li_large_2023,serapio-garcia_personality_2023,dhingra_mind_2023}. These delineated dimensions afford researchers numerous opportunities to enhance the human-likeness of agents\cite{yao_react_2023,hagendorff_machine_2023,driess_palm-e_2023,noauthor_frontiers_nodate,liu_training_2023,fan_minedojo_2022}.

Current research on human-like agents in simulation environments predominantly focuses on their extrinsic aspects. At the level of individual behavior, Danny Driess\cite{driess_palm-e_2023} proposed a multi-modal task-executing agent designed for real-world scenarios. Projects such as MINEDOJO\cite{fan_minedojo_2022} and Voyager\cite{wang_voyager_2023}, leveraging the expansive game world of Minecraft, explore the individual learning behaviors of agents within such virtual settings.

Regarding group-level behaviors, Generative Agents\cite{park_generative_2023} have been developed to enable agents to perform social actions. Additionally, AgentSim offers a comprehensive architecture for simulating standardized agent behaviors, incorporating a module for tool use among agents\cite{lin_agentsims_2023}.

Nowadays, Researchers are gradually realizing the practical computational significance of an Agent's personality on human-like behaviors\cite{ghasem-aghaee_cognitive_2007,aher_using_2023,ma_understanding_2023,ziems_can_2024}. In the study of Humanoid Agents\cite{wang_humanoid_2023}, researchers explored how emotions and needs influence an Agent's planning, largely addressing the overly rational and stable machine-like state exhibited by former Agents\cite{hagendorff_machine_2023}. However, research on embodied agents in virtual environments seldom discusses other intrinsic dimensions. Expanding the scope to other Agent application scenarios, such as conversational interaction Chatbots\cite{tu_characterchat_2023}, we can still find some works focusing on the intrinsic aspects of Agents, such as employing psychological theories, especially cognitive psychology, to test and enhance the cognitive abilities of personality agents, and some works focusing on modelling emotions, especially emotional recognition and empathy\cite{noauthor_emotional_nodate,habibi_empathetic_2023}. In terms of character, some works simulate the static personality and knowledge of historical figures based on external data\cite{noauthor_living_nodate}, and others create chatbots with different static personalities based on personality psychology theories\cite{tu_characterchat_2023}.

The intrinsic personality of real humans is not immutable. In psychology, researchers believe that personality is continuously influenced and changed by external factors\cite{kampmann_feedback_2012}. In existing discussions on research involving embodied agents, we rarely find studies detailing the process of such human-like intrinsic personality evolution or how intrinsic personality evolution affects an agent's extrinsic behavior.
\subsection{Personality Evolution Architecture}
The term "personality" refers to the sets of predictable behaviors by which individuals are recognized and identified\cite{ghasem-aghaee_cognitive_2007,matthews_personality_2003,goldberg_structure_1993,hampson_personality_2012,edwards_measurement_1973}. Personality analysis and description are widely studied in the field of personality psychology, a field that proposes well-known personality classification methods such as the Big Five personality traits\cite{gosling_very_2003} and the Myers-Briggs Type Indicator (MBTI)\cite{coe_mbti_1992}.

Regarding the mechanisms of personality affect behavior, the widely accepted definition of personality refers to cognitive, emotional, and character traits that shape behaviors\cite{zuckerman_psychobiology_1991}. In previous work, researchers believe that personality changes are related to cognitive complexity\cite{ghasem-aghaee_cognitive_2007}. Another study has proposed cognitive conceptual frameworks, including reasoning, retrieval, learning, and grounding\cite{sumers_cognitive_2024}. Further research has summarized the conceptual architecture of agents in more human-like terms of perception, brain, and action\cite{xi_rise_2023}. In research on agents, Some agents can adjust emotional responses and engage in emotional resonance\cite{hasan_sapien_2023}. Additionally, some work has explored character portrayal based on large language models\cite{serapio-garcia_personality_2023}. However, current architectural approaches to agent personality are mostly partial simulations, not yet producing simulations of an integrated personality that can influence behavior.

Moreover, human behavior and personality always evolve in the real world, which presents a substantial challenge for agent simulation. A realistic approach to this changing process is dynamic personality, which considers the variability of personality traits based on changes at the corresponding personality levels. In early research, some works used discontinuous personality update models to achieve the purpose of personality evolution, and some works dynamically simulated personality through the establishment of fuzzy logic agents. These works, trained on traditional models and primarily utilizing a parametric approach to control personality traits, are often constrained by the number of parameters. This limitation results in a lack of realism in agent personality changes. At the same time, different parameter change processes are often directly inferred by the AI model, lacking an interpretive description of the intermediate process, and observers cannot intuitively understand the reasons for character personality changes. As for current work discussing generative model-based agents, they only simulate static personality at a specific time point and have not yet simulated dynamic personality agents that evolve.

\begin{figure*}[h]
  \centering  \includegraphics[width=\textwidth]{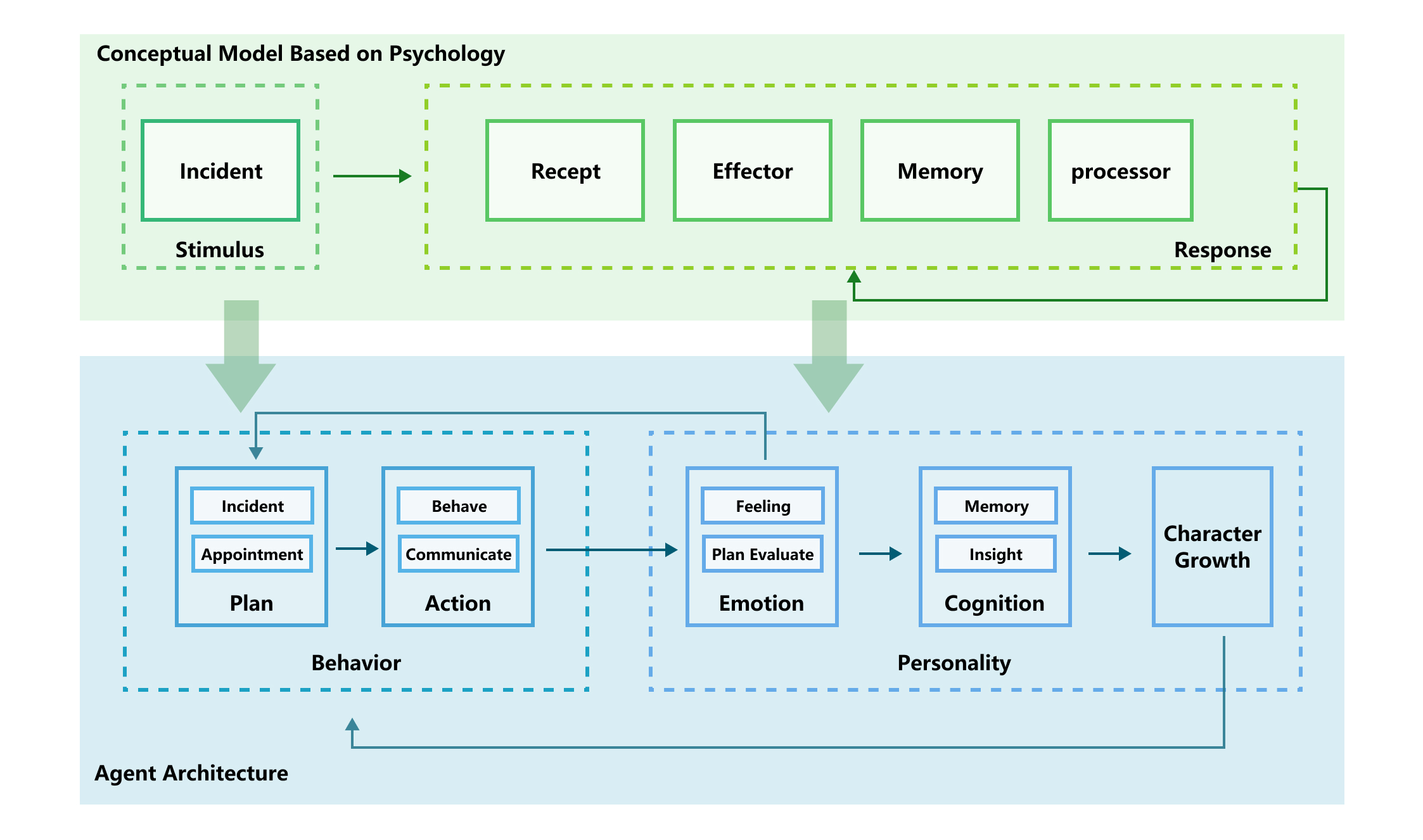} % 确保xxxxx
  \caption{Evolving Agents architecture is inspired by psychology stimulus-response theory, encompasses two subsystems: Behavior and Personality}
  \Description{Evolving Agents Architecture}
\end{figure*}

\section{AGENT ARCHITECTURE}
We first need to understand how human personalities are composed and evolve to enable agents' evolution. The field of psychology already has many rich studies related to this issue\cite{matthews_personality_2003,anderson_embodied_2003,jr_embodiment_2005,solso_cognitive_2005}. Inspired by the stimulus-response theory in behaviorist psychology\cite{bargh_beyond_2000,clark_learning_2018}, we wish for agents to gain stimulation and feedback through interaction with simulation environments, thereby affecting the development of the Agent's personality. 

Based on the psychology theory discussed\cite{noauthor_experimental_nodate,noauthor_classics_nodate,vandenbos_apa_2007,baldwin_mental_1906}, we propose a novel agent architecture - Evolving Agents, which encompasses two subsystems: Behavior and Personality, as shown in figure 2. The Behavior system guides the execution of observable, personalized actions. The personality system establishes a cognitive mechanism with individual characteristics capable of continuously updating personality based on feedback from external information. During simulation, the Behavior and Personality systems continuously influence each other. The planning and behavior of the Agent will take its characteristics into careful consideration. Simultaneously, the experiences generated from an agent's daily actions impact personality through memory and insight. This process of mutual influence forms a feedback loop pattern, facilitating continuous personality evolution.

The behavior system is comprised of two main modules: Plan and Action. Its execution includes personalized planning for the Agent, social engagements and responses, and behavior execution functionalities. The Personality system contains three main modules: Emotion, Cognition, and Character Growth. Its execution involves subjective emotional perception, daily information filtering, memory decay management, comprehensive context-aware idea summarization, and the growth of the Agent's character. Each module is empowered by large language models like GPT-4 when dealing with its input.

\subsection{Agent Behavior}
The Behavior system is responsible for executing planned actions\cite{wang_survey_2024} within a simulated environment. Building on the existing architecture of Agent simulation\cite{caron_identifying_2022,curry_computer_2023,sumers_cognitive_2024}, which includes planning (Plan) and executing plans (Action)\cite{zhou_least--most_2023,song_llm-planner_2023,noauthor_230516291_nodate,shen_hugginggpt_2023}, we introduce three innovative elements in the Plan Module: Characteristic Plan, Goal-based Mechanism, and Specific Post-Process. In the Action Module, we propose a new method to describe the Agent's social relationships and enhance the continuity of conversations between Agents.
\subsubsection{Plan Module}
Characteristic Plan: We aim to reflect agents' personality differences in Evolving Agents' behaviors. Therefore, in the design of the planning module, we consider not only environmental information as inputs but also integrate agents' character and memory as references for GPT-4 to generate plans. This mechanism ensures that each Agent can create a schedule aligned with its personality. The specific mechanism by which the Personality system affects the Plan module will be discussed in section 3.2.
\begin{figure}[h]
  \centering
  \includegraphics[width=\linewidth]{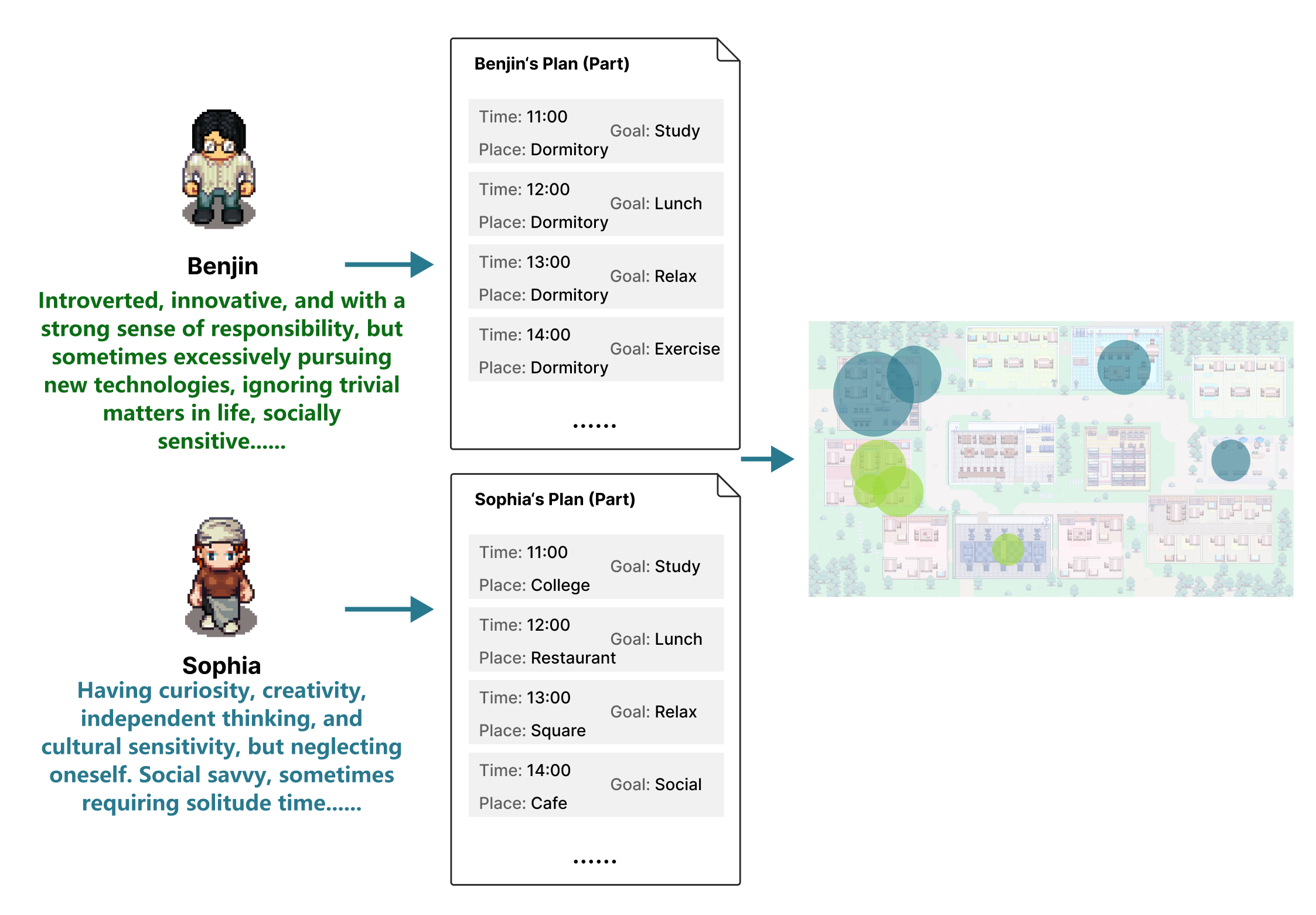}
  \caption{Each Agent can create a schedule that aligns with its character. For example, the enthusiastic and open Isabella prefers to leave her dorm and make social plans compared to the shy Benjamin}
\end{figure}

Goal-based Mechanism: We aim to construct a mechanism that can easily analyze agent behavior and personality variations. Therefore, we referenced relevant perspectives on behavior and motivation from self-determination theory\cite{noauthor_what_nodate}, considering planning as an expression of behavioral motivation. In our specific design, we defined categories of goals (such as learning, relaxation, exercise, etc.) and required agents to understand and refer to these goals while planning, as shown in Figure 3. This ensures that agents consider the motivations behind each planned behavior, guiding their thought process effectively. Such a goal-based mechanism also aids observers in summarizing user behavior patterns, making it easier to analyze related pattern evolutions. We will discuss this in detail in Chapter 5.

Specific Post-Process: We provide different planning goals with unique post-processes. Taking "Appointment" as an example, most existing Agent simulation systems use distance-based and probability-based triggering logic for Agent interactions, as is shown in Figure 4. We aim to enhance the proactivity of Agents in interacting with others, fostering more "conscious" and "prepared" social activities in the simulated world. In the planning phase, Agents can consider their relationships with other Agents and their current state to decide whether to extend social invitations. The invited Agents respond to the invitation, giving reasons for acceptance or rejection. Agents weigh existing social plans against new ones for overlapping appointments, prioritizing them based on personal development benefits. Following this request-response process, both parties adjust their plans based on the outcome.

For example, Sophia wants to meet Isabella at a cafe at 16:00 to discuss, believing that Isabella's appreciation for creative expression aligns with her dream of becoming an inspiring and successful writer. Subsequently, Isabella accepts Sophia's invitation, reasoning that her preference for in-depth conversations over superficial socializing perfectly complements Sophia's personality, suggesting that their meeting could lead to deeper and more meaningful discussions. Interestingly, when faced with simultaneous social invitations, even though Isabella initially planned to engage with another friend, Benjamin, during this time, she considers the conversation with Sophia potentially more crucial for her personal development and thus accepts this invitation. Please refer to the image for details on the process.
\begin{figure}[h]
  \centering \includegraphics[width=\linewidth]{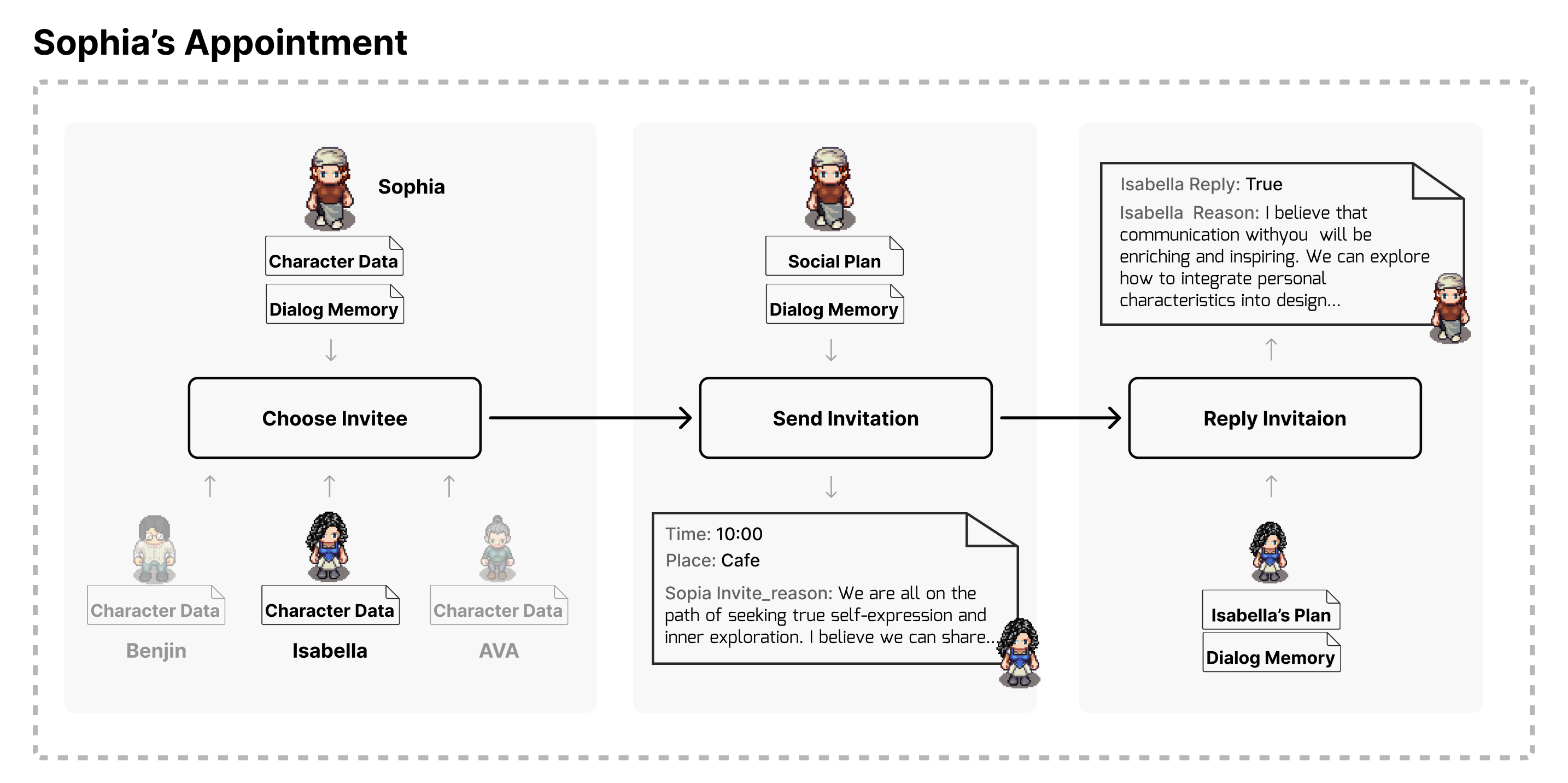} % 确保xxxxx
  \caption{Post process procedure of appointment plan}
  \Description{Evolving Agents architecture}
\end{figure}
\subsubsection{Action Module}
Action Mechanism: When the game time reaches the scheduled plan time, the Agent executes the specific plan. By considering the current plan and personality information, GPT-4-turbo generates a detailed, objective, natural language description of the Agent's actions regarding the plan. Before executing the plan, the Agent must check if the environmental conditions meet the plan's objectives, adopting the item interaction condition logic from Generative Agents\cite{park_generative_2023}. If the target interaction object is occupied, the Agent will replan until conditions are met. Once the environmental object target for executing the plan is secured, the Agent will take over the interaction spot (in social scenarios, reserving adjacent spots for its social peers) and perform the Action. 

Continuously Developing Communication: When two Agents come within a certain distance of each other, there's a probability of triggering a dialogue, depending on their social tendencies and current state (for example, if an Agent's plan is social, it will actively initiate a conversation). From a social psychology perspective, such interactions allow Agents to gain insights from others with different personalities, serving as a vital external stimulus for their personality development. We aim to gradually increase the depth of conversation between agents, enabling them to generate more valuable insights. To achieve this, we have implemented the dialogue memory. Whenever a conversation occurs, the Agents will review the summary of content previously discussed with the current interlocutor. Then, it determines a "further" new topic based on the previously discussed topics and discusses this new topic. This dialogue mechanism enhances the inspiration and viewpoint generation. For example, Isabella and Sophia's discussions deepened from personal development and societal needs to specific methods for maintaining creativity and mental health under social pressure, the procedure of which is shown in Figure 5. This mechanism mirrors real-world communication, where familiarity and discussion depth lead to increasingly inspiring and detailed insights on certain topics.
\begin{figure}[h]
  \centering  \includegraphics[width=\linewidth]{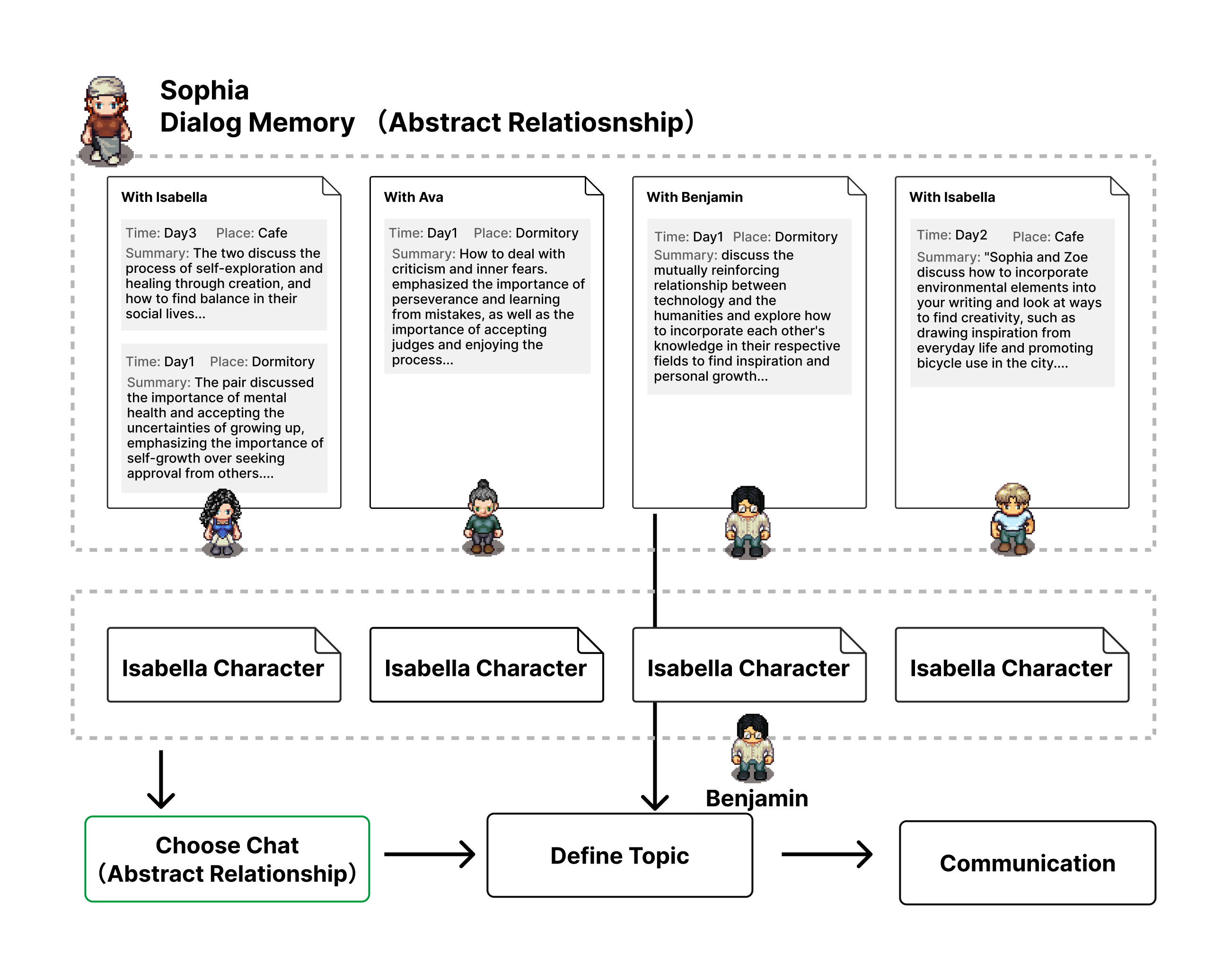} % 确保路径和图片文件名正确
  \caption{Agent chooses their communication partner and defines the topic to talk about based on the dialogue memory}
\end{figure}
Abstract Agent Relationship: Evolving Agents does not use parameters or specific categories to describe agents' relationship with others\cite{wang_humanoid_2023}. We found that doing so would excessively predetermine interactions with certain Agents. We enable large language models to select preferred conversation partners based on agents' dialogue memory. Simultaneously, we require agents to provide reasons for their choices. This approach results in more rational and intriguing social choices, enabling agents to openly choose conversation partners and gain diverse perspectives on their dilemmas. For instance, Isabella, who is balancing external expectations with personal aspirations, might focus on issues like personal dreams and social responsibility with Sophia, a literary writer interested in social insights, or focus on balancing personal pursuits and intimate relationships with Isla, a romantic filmmaker, as is shown in figure 6. This process can help agents explore more possibilities during their Personality Evolution.
\begin{figure}[h]
  \centering   \includegraphics[width=\linewidth]{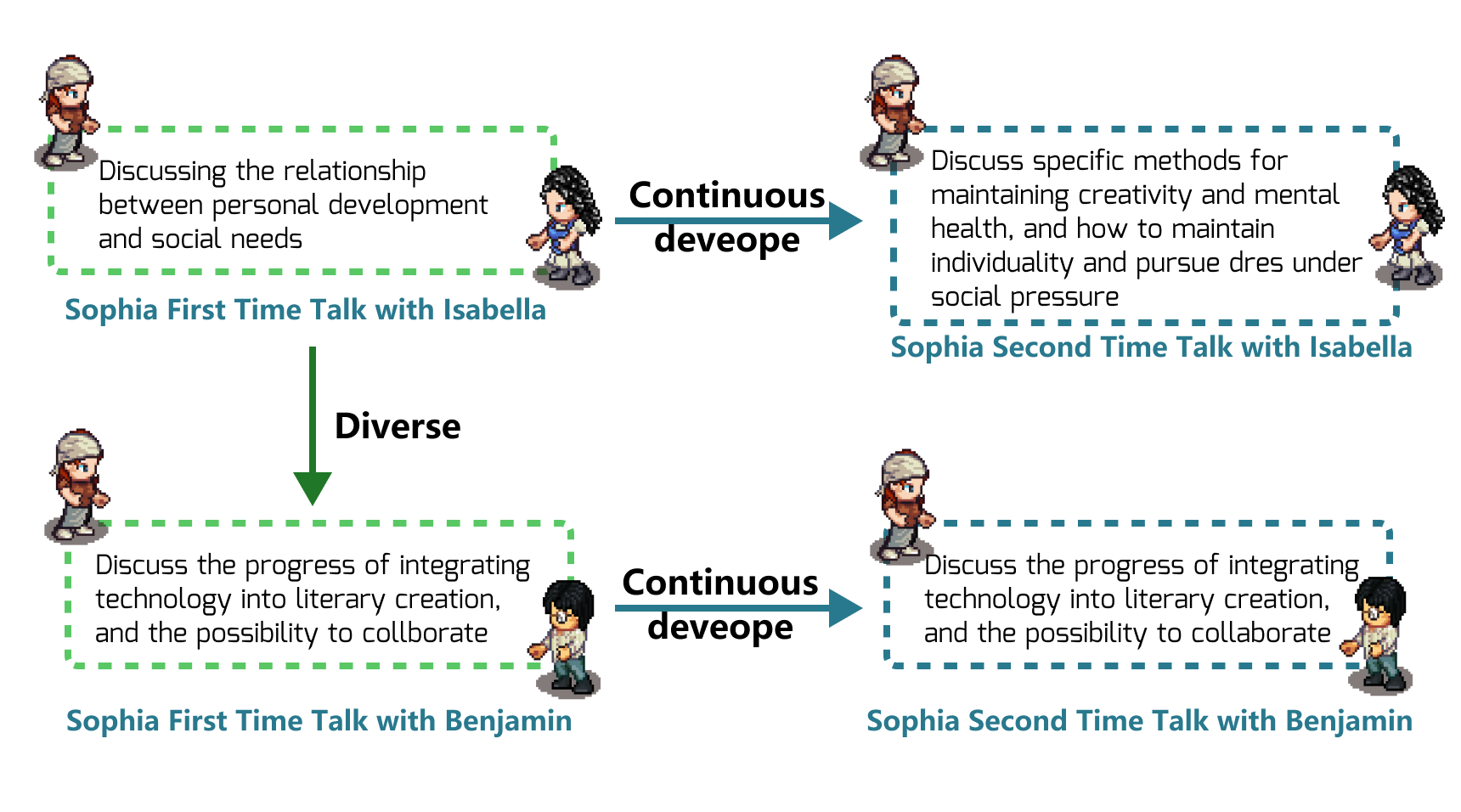}
  \caption{The Agent discusses different topics with various partners, and the depth of each topic continuously increases }
\end{figure}
\subsection{Agent Personality}
In the Agent Behavior System described above, Agents produce a variety of objective behaviors through interactions with their environment and other Agents. From the cognitive psychology perspective, these behaviors can be viewed as stimuli. The Agent Personality system is responsible for processing these stimuli, generating feedback, and thus driving the evolution of the Agent's personality. The widely accepted definition of personality in psychology encompasses cognitive, emotional, and character traits that influence behaviors\cite{zuckerman_psychobiology_1991}. We adapt these elements from psychology theory to fit the operational logic of Agent Architecture, ultimately identifying three main modules within the personality system: Emotion, Cognition, and Character Growth.
\subsubsection{Character Structure}
Five Dimension-Based Character Structure: In psychology, the character is defined as information that includes habits and tendencies in thought and action\cite{snowden_oxford_nodate}, which play a crucial role in their behavior. From the system design perspective, a character directly describes an individual's traits and state. 

A structure that can precisely and comprehensively describe characteristics is essential for effectively integrating and transmitting personality information to each module within an agent architecture. We constructed a five-dimension Character Structure based on psychological theories, including Basic Information, Current State, Traits\cite{noauthor_introduction_nodate}, Conflict\cite{noauthor_apa_nodate}, and Preference\cite{sen_behaviour_1973}, as is shown in figure 7. "Basic Information" encompasses fundamental information about the Agent that does not change, such as gender, age, and professiowetate, we detail the Agent's initial life and psychologic in the Current Stateal state. Personality Traits describes the Agent based on the Big Five personality dimensions. In conflict, we detail contradictions within the Agent's personality, facilitating personality evolution around these conflicts. To better describe Agent's behavior characteristics, in Behavior Preference, we set six-dimension criteria, which are composed of the ultimate goal, long-term goal(derived by an ultimate goal), short-term goal(derived by an ultimate goal), daily routine(based on Agent's overall personality), hobbies and venue preference. In practice, we initialize specific agents' personalities using GPT-4-turbo based on this Five Dimension Character structure.
\begin{figure}[h]
  \centering  \includegraphics[width=\linewidth]{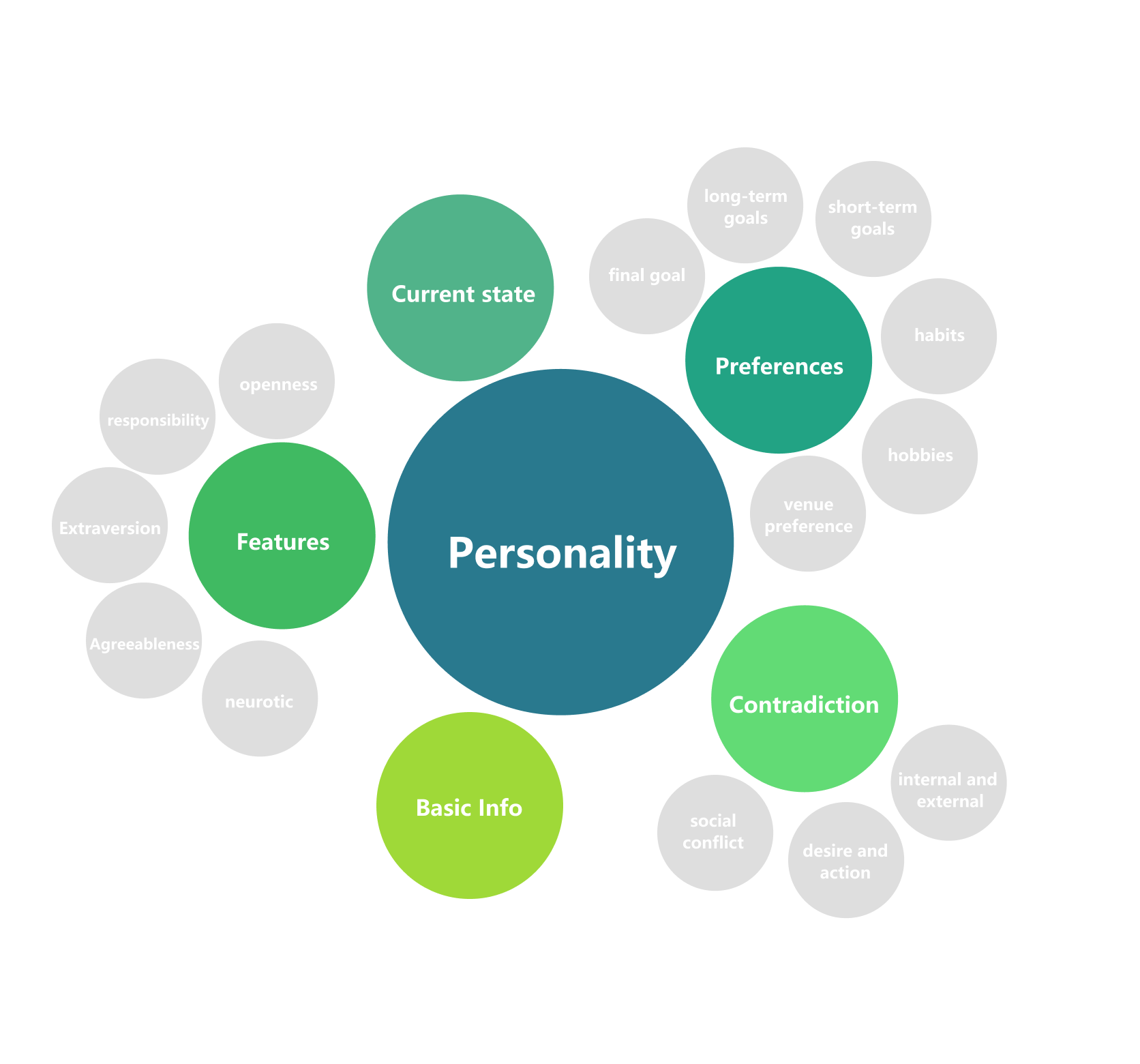}
  \caption{Five Dimension Based Character Structure}
\end{figure}

Two Versions of Character Structure: To enhance the efficiency and accuracy of the system development process, we employ two versions of the Character Structure within the system: a full version and a summary version. The full version includes a complete description, while the summary version condenses and summarizes the full version. In specific modules, we differentiate between utilizing the full version and the summary version based on the variance in information weight. For instance, in the Behavior System's Plan module discussed earlier, we emphasize the information weight of Behavior Preference from the Character Structure. Therefore, we provide a full description of Behavior Preference to GPT-4-turbo but only the summary versions for other dimensions. As for other dimensions, we only provide the summary version. This approach helps the model focus more on the alignment between Plan and Behavior Preference, effectively reducing the generation time for large language models.
\subsubsection{Emotion}
Timely Experience and Feeling: Human feelings towards the same objective behavior are not constant. Specific subjective experiences and emotional responses are generated in response to external stimuli under particular circumstances. In specific circumstances, outworld stimuli affect us to generate emotional behavior and subjective feelings. As a result, our cognitive process, such as attention, thinking patterns and decisions, is influenced, leading us to differentiated perceptions of external information\cite{solso_cognitive_2005}. Therefore, it's crucial to build a system that can not only comprehensively consider the current state and perform emotional perception on ongoing behaviors, but also ensure that the later cognition processes receive information aligned with the character's current emotional state. At the implementation level, we categorize the Agent's emotional state into seven categories. The Emotion module considers the current personality conflicts and other personality information to update emotions based on the behavior being executed. In detail, we request that GPT4 choose the current emotion class with a first-person perspective description of inner feelings. This module allows Agents to generate subjective descriptions that vary and align with their experiences, such as Sophia feeling excited about writing her novel due to anticipation of creating her literary work or feeling anxious and self-doubtful after hearing the technological innovation in the writing process told by Benjamin, as is shown in Figure 8.
\begin{figure}[h]
  \centering
  \includegraphics[width=0.6\linewidth]{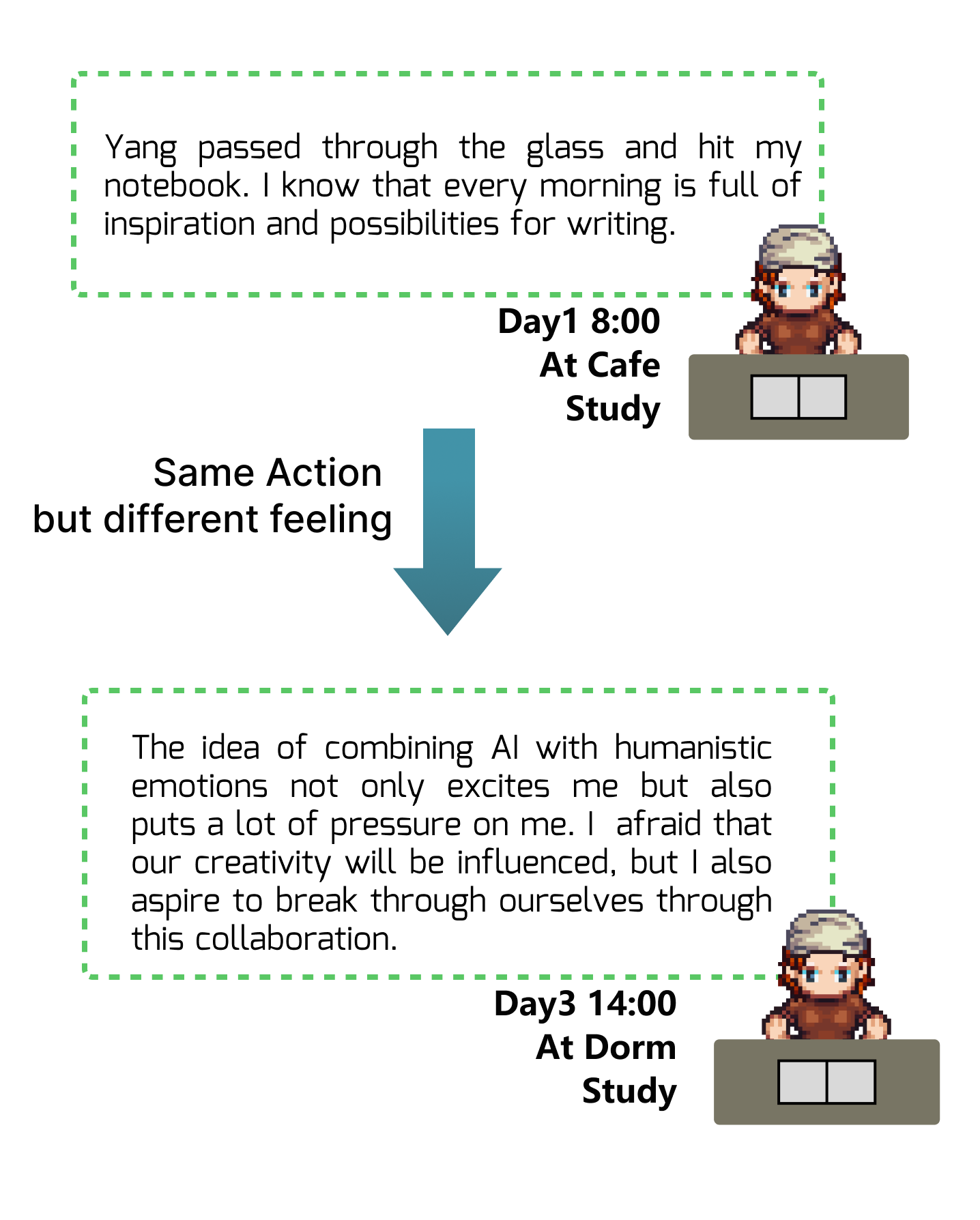}
  \caption{ Agents generate subjective descriptions that vary and align with their experiences}
\end{figure}

Emotion-Based Plan Evaluation: Previous research has noted that humans don't rigidly follow the plans they set for themselves\cite{minsky_society_1988}. In the design of Humanoid Agent architecture, researchers proposed a dual-system structure of emotion and behavior to allow Agents to adapt their plans based on internal feelings constantly.

The Evolving Agents' Emotion module supports continuous evaluation and adjustment in the Behavior system's Plan module. We simplified the adjustment mechanism of Emotion on the Plan, using a more systematic decision-making mechanism to trigger plan adjustments. Emotions are categorized on a scale from negative to positive, mapped within a 0-7 interval. Significant emotional shifts during an action (e.g., from excitement to calm, crossing a 3-interval span) indicate substantial emotional fluctuations, resulting in which agents should reevaluate the forthcoming plans. This straightforward strategy, as shown in Figure 9, demonstrates believable and dynamic planning capabilities. For example, after a conversation with Sophia, Isabella's mood shifted from excitement to fear (due to the discussion regarding social responsibilities caused pressure), leading her to adjust her original plan and go to a square for relaxation to calm her emotions.
\begin{figure}[h]
  \centering
  \includegraphics[width=\linewidth]{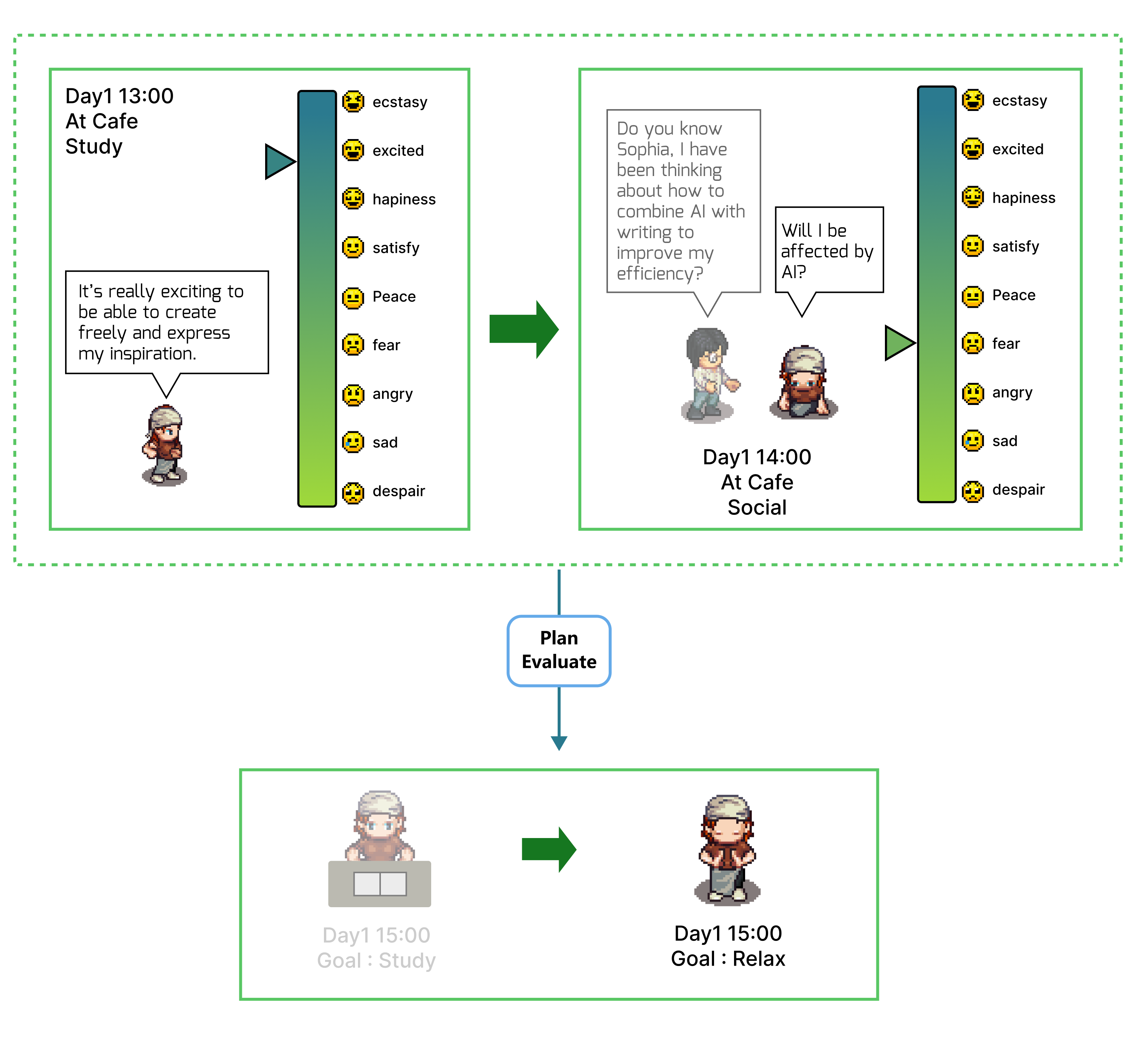} % 确保路径和图片文件名正确
  \caption{Agent evaluates and modifies existing plans based on the emotional fluctuations}
\end{figure}
\subsubsection{Cognition}
Emotion perceives external objective events produced by the Behavior system, generating subjective emotions for each behavior. However, this perception is immediate and fragmented. Every single occurrence and emotion does not directly shape real humans; instead, information is filtered, holistically evaluated and summarized through unique cognitive processes. Instead, humans will filter and evaluate the information we receive. Such a process is Interesting; psychologists also apply computational thinking to understand how humans absorb, comprehend information, and shape their personalities, viewing the human brain as an information processing system that identifies, stores and ultimately reorganizes and reconstructs its system\cite{solso_cognitive_2005}. Based on this concept, we developed the Memory and Insight processes within the Cognition module.

Memory - Characteristic Information Filter: Memory serves as a universal module in the current Agent design, with the Generative Agent architecture introducing the Memory-stream architecture for storing, encoding, and retrieving relevant experiences. Cognitive psychology suggests that people with different personalities tend to notice and acquire information about the objective world differently\cite{minsky_society_1988}\cite{basu_embodied_2004}, indicating the need for a Memory system tailored to individual traits. Therefore, the Evolving Agents place greater emphasis on differentiated information weight assessment. It includes two types of memory data structures: short-term and long-term memory. Short-term memory stores all executed actions and their corresponding emotions as temporary data. Long-term memory, on the other hand, archives information filtered and summarized from short-term memory. The Agent selects from its existing short-term memory based on the personality features within its Character Structure. It describes multiple behaviors and emotions subjectively, storing the outcomes in its long-term memory database. In practice, Agents develop long-term memories that align with personal traits, especially personality contradictions. For example, Benjamin, contemplating interdisciplinary development, pays attention to information outside his field of study, such as reading a literature book in the library. Isabella, grappling with personal versus societal conflicts, takes great interest in the social philosophy discussions provided by Sophia.

Cognitive psychology's memory processing theory\cite{solso_cognitive_2005} suggests that even long-term memories fade over time, understood as the gradual blurring of information. The design of long-term memory also incorporates this fading mechanism by setting a capacity limit for the memory pool. Upon reaching this threshold, the Agent abstracts and condenses the earliest memories to blur them, as shown in Figure 10. 
\begin{figure}[h]
  \centering  \includegraphics[width=\linewidth]{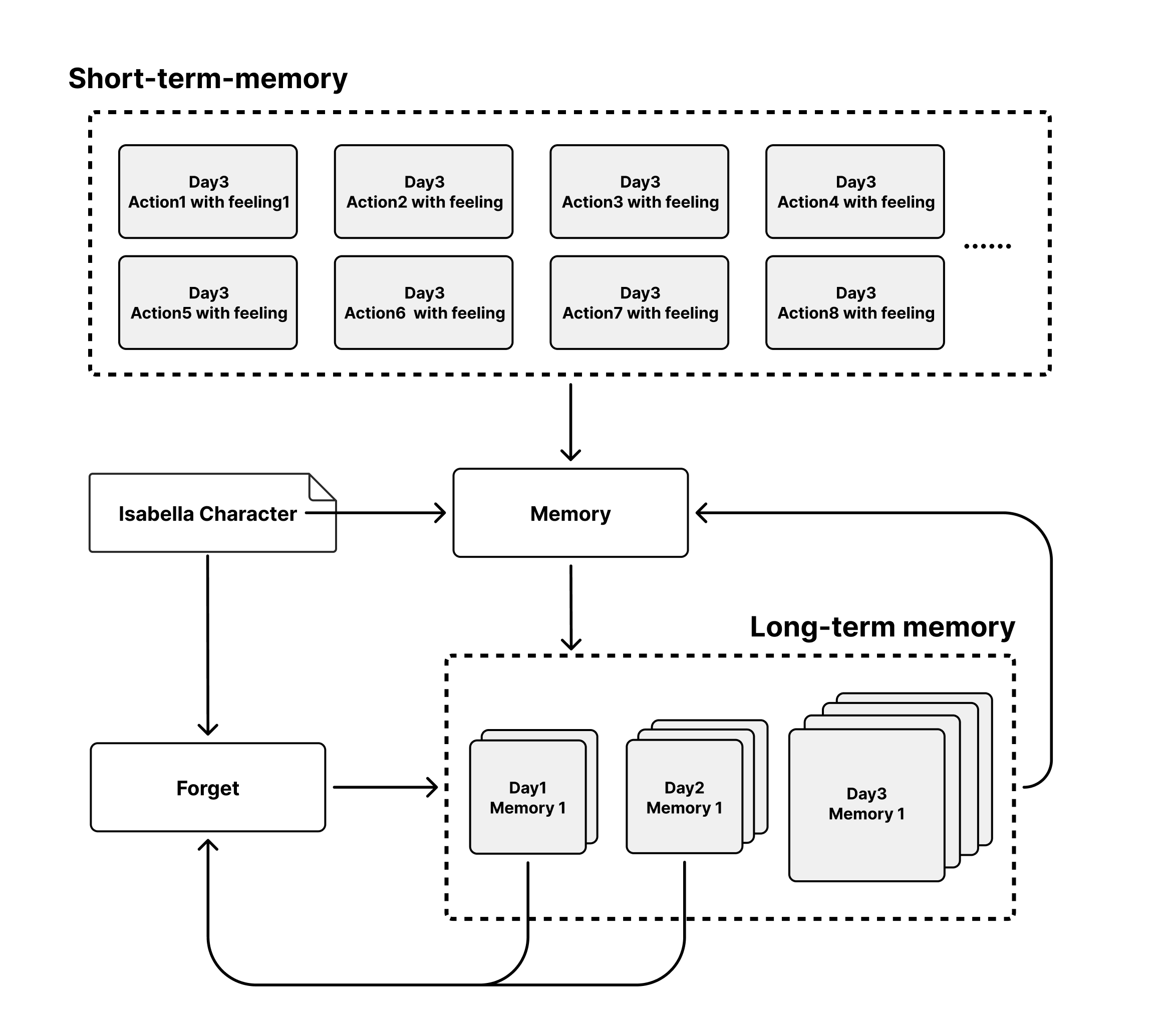} % 确保路径和图片文件名正确
  \caption{The long-term memory mechanism of characters involves extracting information of interest from short-term memory and blurring older information.}
\end{figure}

Insight - Overall Process of the Contextual Incident: Cognitive psychology emphasizes the holism of cognition, where people combine information and relevant contexts to complete an integrated cognitive process\cite{solso_cognitive_2005}. However, in our system, the information filtered through the memory module remains a vast amount of fragmented data, making it difficult for agents to integrate them for holistic thinking. Therefore, we designed the Insight submodule within the Cognition module to help simulate this holistic thinking process. Moreover, we aim for insight to reflect the diversity of individual cognitive processes. Thus, we enable insight to reference the complete character structure, summarizing and reflecting on all events of the day and the long-term memory. In practical tests, we can intuitively perceive that each Agent has unique insights into daily events.

For example, on the first day, Benjamin realizes that becoming a well-rounded individual requires technical breakthroughs and efforts in humanitarian care and social skills. On the second day, Benjamin sees the potential for integrating technology with the humanities, beginning to incorporate a humanitarian perspective into his technical explorations, seeking a harmonious balance between the two. The procedure of such is shown in Figure 11.
\begin{figure}[h]
  \centering
  \includegraphics[width=\linewidth]{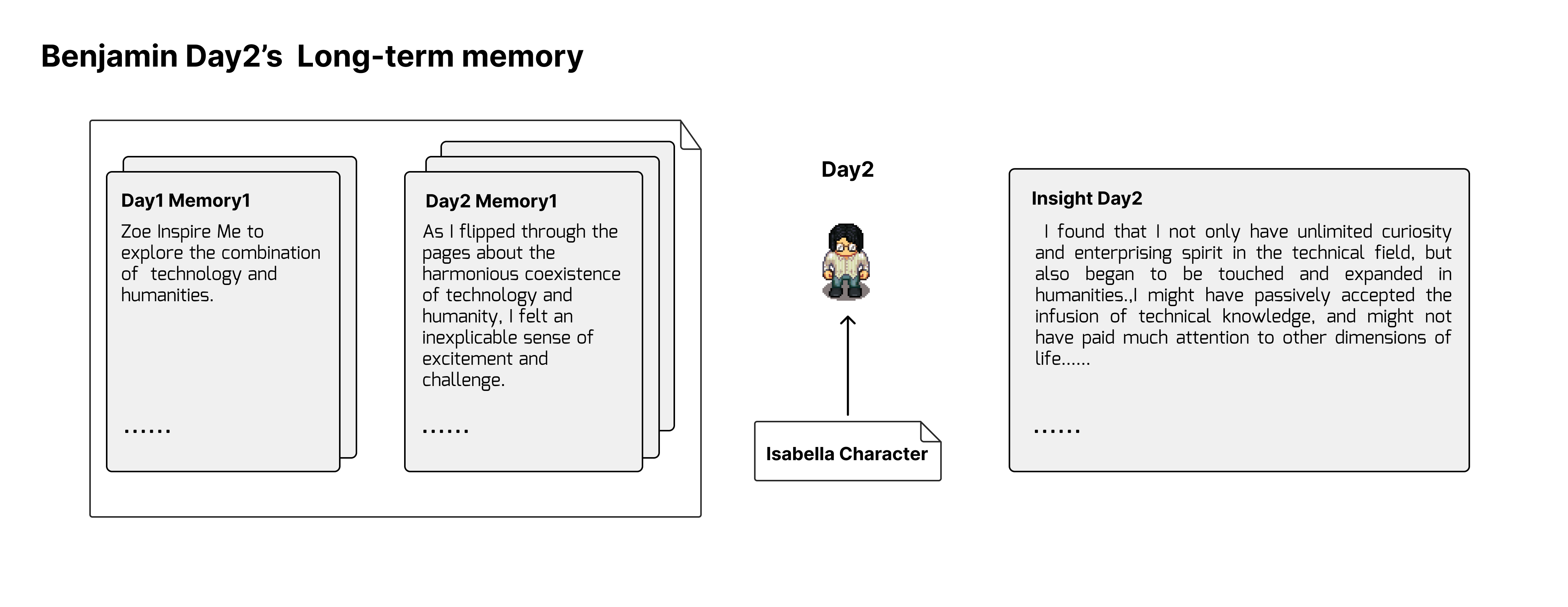} % 确保路径和图片文件名正确
  \caption{Insight enables agents to engage in personalized reflection on daily events.}
\end{figure}

\subsubsection{Character Growth}
Chain-like Update Process: The foundational materials for Agent evolution are prepared after processing external events through the Emotion and Cognition modules. The Character Growth module is responsible for simulating the personality update of agents, transforming daily evolutions into comprehensive descriptions updated within the Agent's Character Structure. This structure doesn't involve complex algorithmic logic but demands high precision in prompt engineering to depict personality changes accurately. We designed the Character Growth process in a chain-like pattern after extensive testing and consideration of the interrelations within the Character Structure, dividing it into four sequential sub-processes based on the Character Structure dimensions, excluding Basic Information: State Modify, Feature Modify, Contradiction Modify, and Preference Modify. In the design of specific prompts, we maintain Character Structure dimensions, continuously summarizing new personality information based on the Agent's current state and filling it into the corresponding dimensions. In the continuous assessment of subsequent development, this growth mechanism has provided us with traceable and observable descriptions of personality changes.

For example, compared to Day 1's Benjamin, who was immersed in technology, his character now includes the realization that life encompasses more than work or academics, emphasizing health and happiness maintenance. This change was influenced by Zoe's encouragement for a healthy lifestyle during their conversation, as is shown in Figure 12, highlighting Benjamin's personality transformation. Similarly, Isabella's interactions with Sophia and her solitary reflections in the square prompt her to shift from a personality of conformity to one that is more free and authentic.

\section{Simulation Platform}
Like the sandbox simulation environment proposed in the former work, we create a Sandbox using the Unity engine, as shown in Figure 12. In this sandbox, Agents will move and exhibit their actions based on the computational results of Evolving Agents architecture, making it easier for observers to interact.
\begin{figure*}[h]
  \centering
  \includegraphics[width=\textwidth]{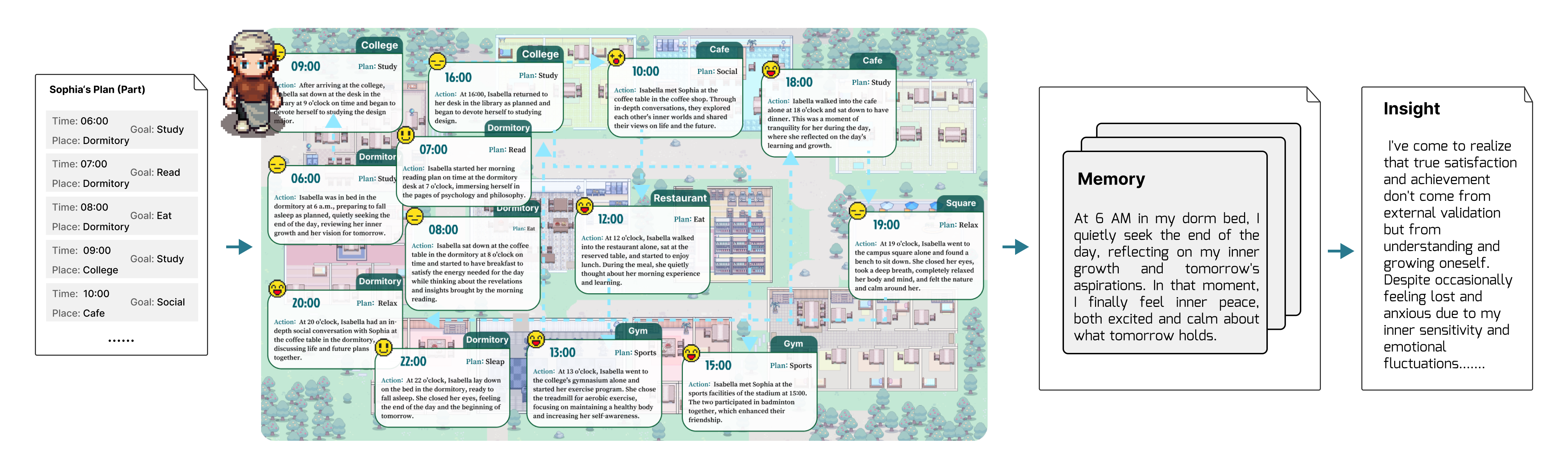}
  \caption{In the sandbox world, users can observe the Agent's daily behavioural activities and inner thoughts.}
\end{figure*}
\subsection{Building System}
\subsubsection{Three Level Based Environment}
Describing information contained in the environment to the large language model is vital. Agents need to be aware of the objects in the environment, whether they are interactable and how to interact. When describing information contained in the environment, we offer a novel thought. We designed a building-place-goal three-level data structure, as shown in Figure 15, to describe the environment. This structure enables us to efficiently convey structured environmental information to large language models, thus allowing agents to fully consider existing environmental information and locations during their planning phase and improving their efficiency in understanding the environment. This directly manifests that large language models will aim to align their planning objectives as closely as possible with the existing environmental functions and descriptions.
\begin{figure}[h]
  \centering
  \includegraphics[width=0.6\linewidth]{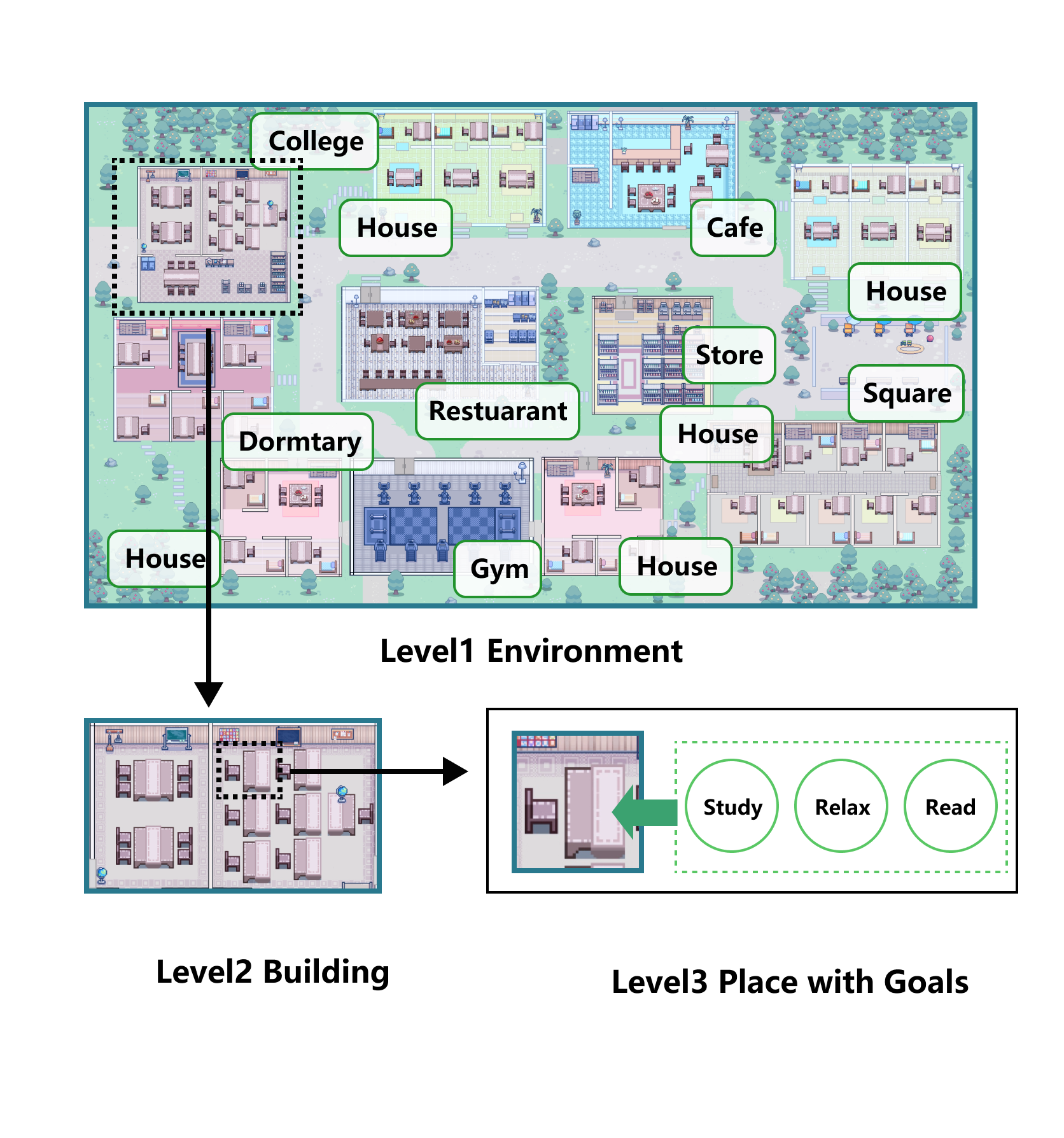}
  \caption{Environment data structure}
\end{figure}
\subsubsection{NPC Action Execute}
The daily plans generated by the Agent planning module are read and translated into ScriptableObject-type task description files in Unity. When the game time reaches the next plan, the engine will drive the Agent to move to the target location using the A-star algorithm and, upon arrival, play the corresponding animation. To enhance the decoupling of work between the agent architecture and the sandbox, we have detailed programmatic descriptions of the operational rules in the sandbox, such as the Agent's moving speed and spatial information descriptions. This way, the agent architecture can directly generate Agent performance data that fits the logic of the sandbox (for example, if the Sandbox screen running time from the dormitory to the college is 15 minutes, then the agent architecture will consider this 15-minute moving time in advance when executing). This approach ensures the decoupling level during the team development process and enables the users to design and iterate environment settings when art resources for the sandbox are unavailable.
\subsection{Using System}
\subsubsection{Environment Editing}
Inspired by behavioral psychology, Evolving Agents is also envisioned as a research tool for the field. In behavioral psychology, external stimuli are one of the primary variables in personality development or change. In this potential demand scenario, we hope the system will allow users to easily modify the system's environmental settings to meet their future needs. Environmental updates in the game interface require corresponding art resources, but expecting users to provide these resources is impractical. However, as mentioned in section 4.2, agent architecture and sandbox are highly decoupled. Users can edit the environment and obtain simulation results without waiting for the art scenes to update. We have developed a tabular operation experience for environmental editing to lower the learning threshold for user editing. Users can directly control environmental variables by adjusting an online CSV table, thereby achieving adjustments to the Agent's living space.
\begin{figure}[h]
  \centering
  \includegraphics[width=0.6\linewidth]{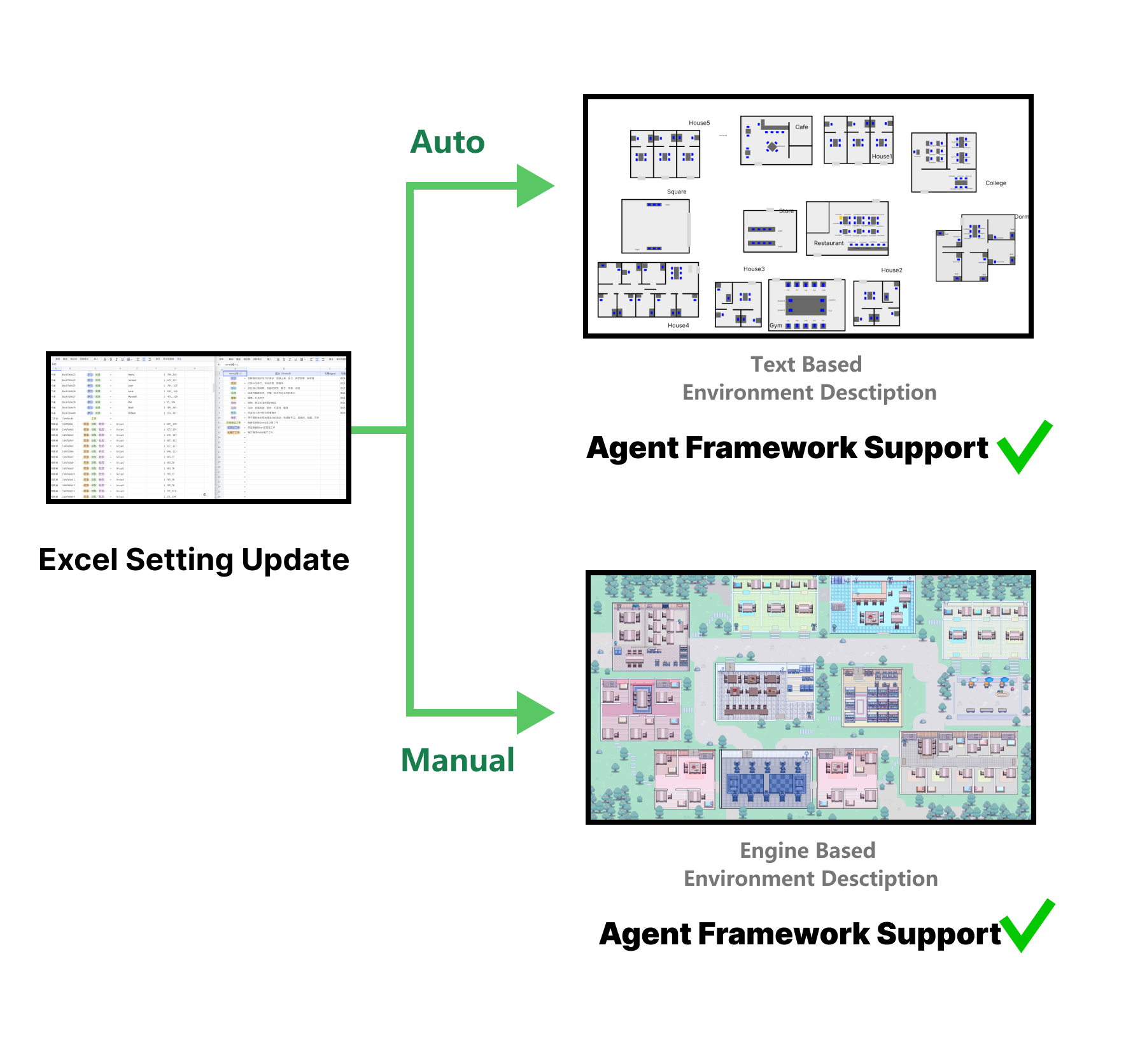}
  \caption{Users can directly update environmental settings by adjusting an online CSV table}
\end{figure}
\subsubsection{Chat With Agent}
Although the current research theme does not encourage users to provide artificial stimuli beyond the simulation process for Agents, we also build relevant user interaction experiences in the Simulative System. We aim for user interactions to be consistent with other Agents from an Agent's perspective. Therefore, users can choose to initiate a conversation with any Agent not currently engaged in a dialogue during the system's operation. This conversation will also be recorded and summarized in the Agent's Dialog-memory, contributing to the Agent's cognitive process.
\section{OBJECTIVE DATA VERIFICATION}
The goal of Evolving Agents is to enable them to perform embodied behaviors according to differentiated personalities in the interactive simulation system and to evolve their personalities continuously based on external information during interaction processes. Around this goal, we propose three more specific research hypotheses:
\begin{itemize}
    \item Evolving Agents can construct agents with perceivable and believable differentiated personality traits.
    \item Evolving Agents' personalities can undergo perceivable and believable evolution during interactive simulation processes.
    \item The personality evolution of Evolving Agents can lead to perceivable and believable changes in their behavior patterns. 
\end{itemize}
Evaluating believability and permeability mentioned above relies on human users' observation and subjective assessment. However, before recruiting users for experimentation, we can first analyze the operational logs of the Agents to verify changes in personality and behavior patterns that have occurred during the simulation from a machine metrics standpoint (like alpha and beta testing in software development processes).

We implemented two types of objective data validations, using the Big Five personality assessment and behavior difference analysis, respectively. By comparing with the ablation architecture, we objectively verified that the Evolving Agents exhibited personality evolution and corresponding behavioral changes during the simulation process.
\subsection{Big Five Personality Traits Assessment}
We aim for Agents to define their current personality based on objective scale questionnaires throughout the simulation process and find the objective evidence of personality changes in Evolving Agents by comparing results over time. The character structure of Evolving Agents is represented based on the Big Five personality traits. Therefore, we refer to existing Big Five personality analysis questionnaires and equip Evolving Agents with the ability to fill out questionnaires, enabling them to continuously undertake Big Five Personality Inventory (BFI-44) assessments\cite{john_big_1999}and to output the analysis results of the Big Five personality traits during the simulation process. In the specific implementation of enabling Evolving Agents to fill out questionnaires, we discover that even when we constrain the Agent to the same state, the results of each BFI-44 assessment vary significantly. The large language models lack a clear standard for reference and comparison. As a result, we design a process for filling out the same set of questions across multiple days in parallel, asking the model to focus on the differences from day to day for comprehensive questionnaire completion, which effectively resolves the issue of lacking a clear standard and reference for questionnaire completion across sessions.

Contrasted with the complete Evolving Agents architecture, we have an ablated architecture that lacks the insight and Character Growth components of the Personality system. This ablation is because this version of the Evolving Agents architecture is somewhat similar to previous Agent architectures, including the basic processes of Plan-Action-Memory-Reflect. Hence, it has a certain referential value.

We use three Agents, Isabella, Benjamin, and Sophia, for a seven-day simulation within a sandbox\cite{park_generative_2023}, calculating their daily scores across the five dimensions of the Big Five personality traits, as is shown in Figure 16 and Figure 17, with the scoring method referenced. Finally, we determined the difference in values between each dimension over two days for each person. \newcommand{\diff}{\Delta s}
Let \[ D = \{ \text{BFI\_EXT, BFI\_AGR, BFI\_CON, BAI\_NEU, BAI\_OPEN} \} \] be the set of dimensions evaluated.
For each dimension $d \in D$, given a series of scores $s_d$ defined as shown below
\begin{equation}
s_d = \{s_{d1}, s_{d2}, \ldots, s_{dn}\},
\end{equation}
% 平均两天差异度
where $n$ is the number of days,  we assess the overall personality change activity$\Delta_{\text{overall}}$ for each person using the following formula:
\begin{equation}
\Delta_{\text{overall}} = \frac{1}{D(n-1)} \sum_{d=1}^{D} \sum_{i=1}^{n-1} \left| s_{di} - s_{d,i+1} \right|
\end{equation}

The Big Five personality change states of Isabella, Benjamin, and Sophia, who have different initial personality settings over seven days, are shown below
\begin{figure}[h]
  \centering  \includegraphics[width=\linewidth]{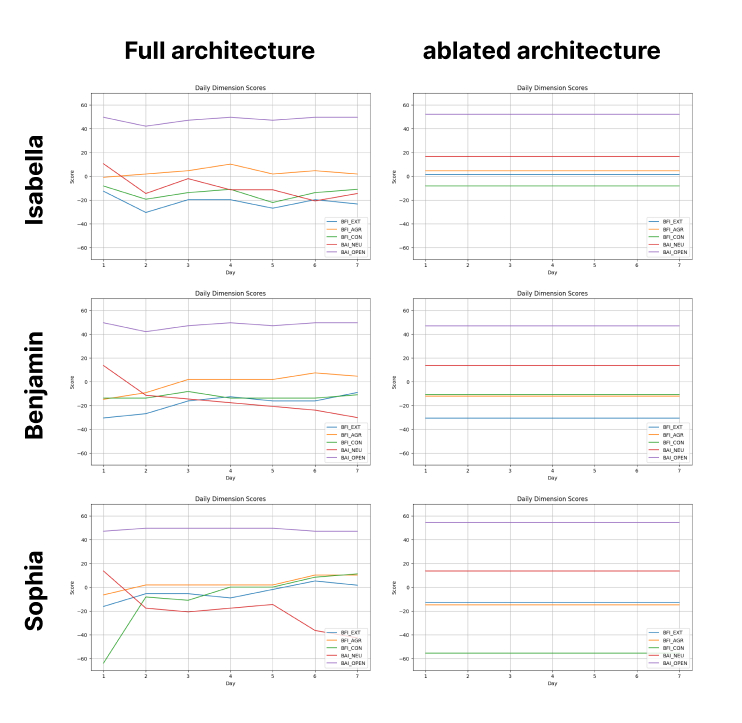} % 确保路径和图片文件名正确
  \caption{Results of the Big Five personality traits assessment}
  \Description{Results of the Big Five personality traits scale assessment}
\end{figure}

We can see that the Agents in the ablated group show no change in their personality assessment results over the seven days, with an average personality difference of zero every two days for all three Agents. In contrast, Agents in the Full Architecture exhibit active personality change states. For Benjamin, the $\Delta_{\text{overall}}$ is 4.37; for Isabella, it is 6.52; and for Sophia, it is 6.77. This data supports our claim to a great extent that the Evolving Agents can lead to personality evolution in Agents during the simulation process.

From the trends in changes across various dimensions, we also discover that Agents with different personalities exhibit distinct patterns of personality evolution. For example, Benjamin, an introverted and stable computer science student, is slowly adjusting his personality toward his target. In contrast, Sophia, a more creative and active writer, experiences rapid changes in neuroticism and extraversion, which intuitively aligns with our understanding of their individual personalities.

However, while displaying divergent changes, the three individuals also show similar development directions. Overall, all three have shown a decrease in neuroticism, possibly due to the positive bias in the content of large language models. This makes it easier for agents to evolve towards more self-consistent and peaceful personality directions.

\subsection{Behavioral Similarity }
In the Agent architecture, we limit the goal selection in the planning phase to a finite number of abstract categories. This allows us to conveniently measure the similarity of daily plans by encoding each day's plan and analyzing them based on the Euclidean distance, thereby providing a measure of behavioral differences between days. 
Given a dataset representing daily goal counts, the Euclidean distance between the behavior on two different days, day $i$ and day $j$, is defined by the formula:
\begin{equation}
E(i, j) = \sqrt{\sum_{k=1}^{N} (x_{ik} - x_{jk})^2}
\end{equation}
where $E(i, j)$ represents the Euclidean distance between day $i$ and day $j$, $x_{ik}$ is the count of goal $k$ on day $i$, and $x_{jk}$ is the count of the same goal on day $j$. $N$ is the total number of distinct goals observed over all days. This calculation assumes that for any goal not observed on a given day, its count is zero.

Given a distance matrix $D$ of size $n \times n$, where $D_{ij}$ represents the Euclidean distance between the behavior on days $i$ and $j$, the overall activity level, denoted as $A$, is calculated as the average of all upper triangular elements of $D$ excluding the diagonal. The formula gives this overall activity level:
\begin{equation}
A = \frac{\sum_{i=1}^{n-1} \sum_{j=i+1}^{n} D_{ij}}{\frac{n(n-1)}{2}}
\end{equation}
where $n$ is the total number of days. The denominator, $\frac{n(n-1)}{2}$, represents the count of elements in the upper triangular part of $D$ excluding the diagonal, effectively counting each pair of days once.

We use the same ablation control group as in the Big Five personality trait analysis for the same reason. This is the version of Evolving Agents most similar in architectural description to past studies; hence it holds referential value.

We use three Agents, Isabella, Benjamin, and Sophia, for a seven-day simulation within a Sandbox, calculating the behavioral differences between days. The analysis results of behavioral similarity are as follows:
\begin{figure}[h]
  \centering
  \includegraphics[width=\linewidth]{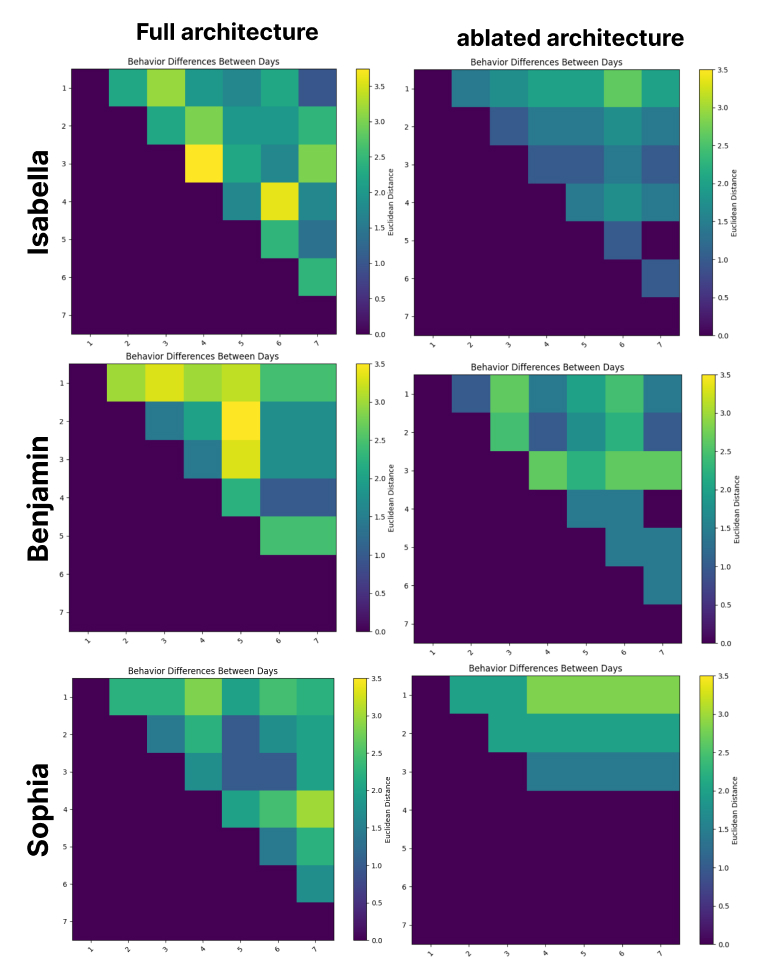} % 确保路径和图片文件名正确
  \caption{Results of the Behavioral similarity assessment}
  \Description{Results of the Behavioral similarity assessment}
\end{figure}
 For Isabella, the activity level in the full architecture is 2.293, while the overall activity level in the ablated architecture is 1.416; for Benjamin, the overall activity level in the full architecture is 2.152, while in the ablated architecture, it is 1.742; for Sophia, the overall activity level in the full architecture is 1.949, while in the ablated architecture it is 1.474. It is evident that compared with the ablated Agent architecture, Agents in the full architecture exhibited more active behavioral changes.

When specifically planning behavior in the Plan module, there are no differences in the input data composition between the full architecture and the ablated architecture of Agent planning. The only difference is that the full architecture agents receive evolved personalities daily, whereas the ablated architecture repeats the initial personality settings. This also indirectly verifies that personality differences are one reason for changes in behavior patterns.

\section{HUMAN EVALUATION}
After validating the objective data, we prepared and conducted a human evaluation experiment, continuing with the research hypotheses mentioned earlier. For Hypothesis 1, we asked human evaluators whether the personality differences between different Agents are perceptible and believable. Regarding Hypotheses 2 and 3, we asked human evaluators whether the personality differences between different Agents are perceptible and believable to the users. 
\subsection{Evaluation Procedure}
We recruited 30 human evaluators (14 male, 16 female) who signed informed consent forms to participate in our study. Their ages ranged from 18 to 34 years, encompassing various disciplines and professions, with good reading and analytical skills, and all having a bachelor's degree or higher education.

Before the formal experiment, we conducted a preliminary study. In this pre-experiment, we first selected a 10-day log of one of the agents. We presented it through front-end animations on the simulation platform for two human evaluators to assess. During the pre-experiment, we identified the following issues: 1) In practice, human evaluators were overly distracted by images and other irrelevant information, making it difficult for them to focus on the Agent's personality growth and behavioral changes. 2) The amount of textual information from 10 days of evolution was too overwhelming for general evaluators, making it difficult for them to fully understand the complete process of the Agent's changes over the ten days; 3) The evolution of the Agent's personality and behavior is a highly abstract and generalized simulation of human personality evolution. Observing an agent for a long simulation time is time-consuming for human evaluators. Human evaluators must detach from the normal human perception of time and focus on evaluating the evolution process.

Therefore, we used text records of an agent's logs over three consecutive days for user assessment in the formal experiment. Additionally, before the experiment, we emphasized to the human evaluators the importance of reducing consideration of the time factor and focusing on the believability of the evolutionary process itself.

In the formal experiment, we first asked human evaluators to spend about an hour reading and taking notes on the logs of three agents and four ablated versions of one Agent. We also asked them to use the think-aloud method to communicate their thoughts to the experimenters during the reading process. After completing the reading, we asked them to complete questionnaires for the control and ablation experiments based on their notes. During this process, they could continually refer to the reading materials to ensure the accuracy of their choices. The questionnaires we set up focused on two main aspects: user believability rating and ablation control experiments.
\subsection{User Believability Rating}
In the user believability rating, we posed two questions regarding the human-like effects of Evolving Agents: "How similar are the personality and behavioral differences among the three groups of agents to those between real humans?" and "How similar is the growth trajectory of personality and behavior of the three groups of agents to that of real humans?" We asked the human evaluators to score these two questions on a scale of 0 to 5. 

In the user believability rating, the scores for the human-likeness of personality and behavioral differences of Evolving Agents, after a Wilcoxon signed-rank test, yielded a statistic of T = 11.0, p < 0.001. This very small p-value indicates that the assessed personality differences are statistically significant compared to a neutral rating (3). This can be interpreted to mean that the scores in the survey were generally above the median, with participants broadly considering the personality differences to be significant and reasonable. This effectively supports Hypothesis 1: Evolving Agents can construct agents with perceptible and believable differentiated personality traits.
\begin{figure}[h]
  \centering
  \includegraphics[width=0.8\linewidth]{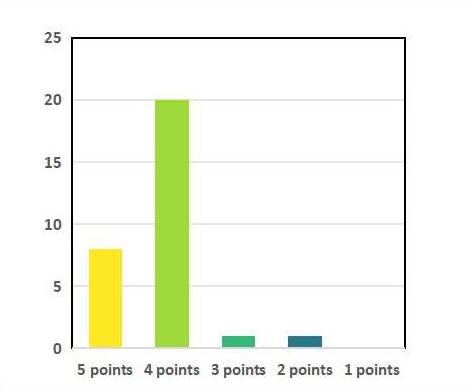}
  \caption{believability of personality and behavioral differences of Evolving Agents}
  \Description{human-likeness of personality and behavioral differences of Evolving Agents}
\end{figure}
For the scores on the human-likeness of the growth trajectory of personality and behavior of Evolving Agents, after a Wilcoxon signed-rank test, the statistic was T = 11.5, p < 0.001. The very low p-value indicates that the null hypothesis can be rejected, suggesting there is sufficient evidence to claim that the growth trajectory and personality development scores are statistically significantly higher than the median level (3), meaning participants generally found the growth trajectory and personality development to be obvious and reasonable within the assessed context. This effectively supports Hypotheses 2 and 3: In interactive simulations, the personalities of Evolving Agents can undergo perceptible and believable evolution.
\begin{figure}[h]
  \centering
  \includegraphics[width=0.8\linewidth]{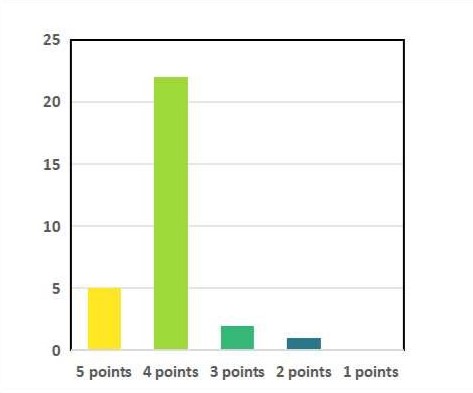}
  \caption{believability of the growth trajectory of personality and behavior of Evolving Agents}
  \Description{human-likeness of the growth trajectory of personality and behavior of Evolving Agents}
\end{figure}

\begin{figure}[h]
  \centering
  \includegraphics[width=\linewidth]{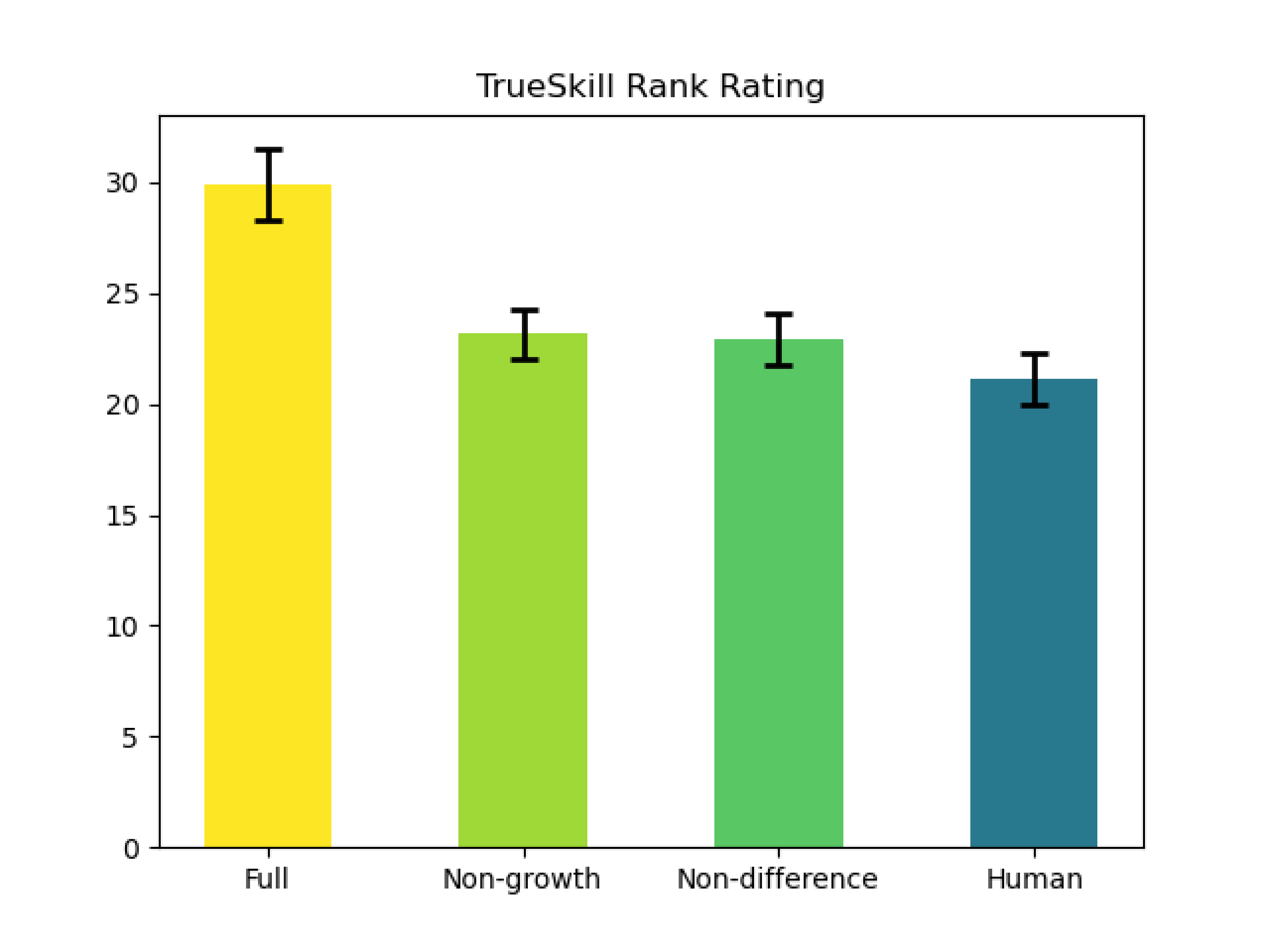}
  \caption{Result of ablation experiment }
\end{figure}
\subsection{Ablation Experiment}
To validate the positive impact of various modules within the Evolving Agents on personality differentiation and personality evolution in terms of perceptibility and believability, we conducted two degrees of ablation studies on the architecture of the Evolving Agent. Compared to the full Evolving Agents architecture, Group 2 had the Cognitive Feelings, Insight, and Character Growth parts of the Personality system removed, consistent with section 5.1. At the same time, Group 3 was further simplified from Group 2 by simplifying the Character Structure, adopting a character description method similar to those used in other existing embodied agent sandbox simulation systems. We asked human evaluators to rank the human-likeness of the growth trajectories of personality and behavior for four experimental groups and to describe the reasons for their rankings in words.

We obtained 30 sets of ranking data from user studies in the ablation experiment. Based on this data, we calculated the TrueSkill ranking scores for the four experimental groups and computed Cohen's d values to quantify the strength of differences between groups. In the TrueSkill scoring system, the $\mu$ value represents the average performance level of an experimental group, while the $\sigma$ value measures the uncertainty of the scores. Additionally, to explore the statistical significance of these results, we conducted a Kruskal-Wallis test, a non-parametric alternative to one-way ANOVA, to confirm the overall statistical significance of differences under different experimental groups. To pinpoint the exact sources of differences, namely which specific group comparisons showed significant differences, we used the Dunn test for pairwise comparisons between experimental groups, and p-values were adjusted using the Holm-Bonferroni method. The experimental results are illustrated in a bar chart, with the height of the bars corresponding to the $\mu$ \ value and the top of the bars marked with error lines, indicating the $\sigma$ \ value.

The final Result shows, Group 1 achieved the best scoring results ($\mu$=29.89; $\sigma$=1.61), followed by Group 2 ($\mu$=23.18; $\sigma$=1.13), and then Group 3 ($\mu$=22.95; $\sigma$=1.15). Comparing the highest-scoring Group 1 with the composite scores of the other three groups, Cohen's d value of 5.33 indicates that Group 1 performed significantly better than Groups 2, 3, and 4 combined. The Dunn posthoc test, both uncorrected and corrected using the Holm-Bonferroni method, showed that the difference between Group 1 and all other groups (Groups 2, 3, and 4) remained significant. At the same time, there was no significant difference among Groups 2, 3, and 4.

Among all experimental control groups, Group 4 performed the worst ($\mu$=21.16; $\sigma$=1.18 ). Surprisingly, users perceived the human group to have the lowest level of anthropomorphism. Following the experiment, feedback from users based on the lowest scores of Group 4 suggested that they found the behavior and personality change process of real humans too stable, with barely any changes observable. This points to a potential issue in experimental design: real humans, indeed do not exhibit many behavioral or personality changes over short periods, and our routines might be somewhat mechanical and dull in a sense. However, overall, the comparison of the first three groups in the ablation experiment, combined with the user credibility rating, effectively supports the three research hypotheses focused on in the study: 1) Evolving Agents can construct agents with perceptible and believable differentiated personality traits; 2) In the interactive simulation process, the personalities of Evolving Agents can undergo perceptible and believable evolution; 3) The personality evolution of Evolving Agents can influence their behavior patterns to undergo perceptible and believable changes, and the ablation experiment method confirmed that Evolving Agents plays a key role in supporting these research hypotheses.
\section{SUPPLEMENT INSIGHT WORKSHOP}
During the inspiration experiment phase, we primarily evaluated Evolving Agents' inspirational effects and limitations as design probes on designers' thinking and design processes. We recruited nine users from the design field, aged between 22-50 years, covering multiple design directions such as architectural design, industrial design, interaction design, game design, etc., all with at least a bachelor's degree.

In the experiment, we first had the designer participants spend 30 minutes observing the daily logs of personality and behavior for three agents, discussing them, and writing down design insights generated from them on paper. We then facilitated a brainstorming session where each designer discussed and associated their own and others' insights, continuously adding new insights.

Lastly, we asked each designer to step forward to present their summary of views and fill out a five-point scale questionnaire on the inspirational effect of Evolving Agents as design probes. The scoring results are shown in the following graph.

Most designers were surprised by such a system design and provided us with many innovative ideas on how such embodied agent simulation systems could be integrated with design processes. For example, a discussant with a background in environmental design believed that this system could simulate users' experiences and feedback in certain spatial environments, thus obtaining firsthand design opinions. A student specializing in drama and film thought it could be possible to draw inspiration from the Agents' lives in a sandbox world to aid in their story creation. In contrast, a student of service design believed this system could help explore new service scenarios tailored to specific groups.
\begin{figure}
  \centering
  \includegraphics[width=0.8\linewidth]{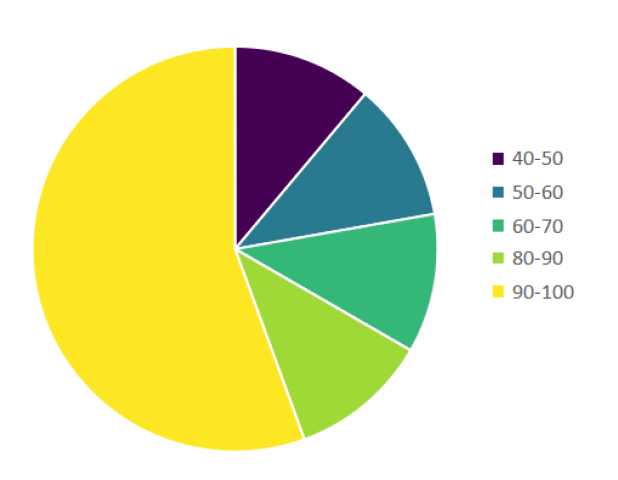}
  \caption{Evolving agents score as heuristic effects of designing probes }
\end{figure}
\section{DISCUSSION}
\subsection{Interdisciplinary Perspectives on Agent Architecture Construction Thinking}
Currently, many simulation Agent architectures have relatively complete capabilities in behavior, learning, and interaction. The architecture of these architectures essentially continues the traditional design philosophy of agents, built around a machine architecture concept that encompasses input, processing, and output from Perception to Mind-like thinking and then to Action. With the development of large language models to this day, their emergent general abilities and intent recognition greatly facilitate interdisciplinary theoretical transfer for researchers designing AI systems or Human-Agent Interaction (HAI), focusing on "how to create embodied agents with diverse and dynamic personalities."

In the era of natural language models, AI can exhibit stronger transfer abilities and general capabilities and can interact with humans through natural language. This characteristic endows natural language models with the ability to "think about problems through natural language." This feature allows us to imitate the entire human cognitive process, simulating human thinking and output in various modules in a natural language manner.

Based on such technological transformation, we believe the current approach to constructing agent architectures can be divided into two steps. First, map the psychology analysis architecture for such human behaviors into computationally feasible processes and architectures. Secondly, in the specific module design, based on relevant psychological theories and combined with human analysis and processing experience, write and debug prompt structures that the natural language models can understand. This ensures that the entire process and specific processing steps are more human-like for large language models.
\subsection{The Potential and Limitations of Embodied Agents' Applications}
\subsubsection{The Prospects of Embodied Agents in Sandbox Simulation Systems}
During the actual experiments, many human evaluators are confused, feeling that the experiences from all four experimental groups are quite human-like, making it difficult to distinguish which one is a real human clearly. Surprisingly, the performance of real humans is the lowest (possibly because our human actor has a stable mood and a rather mundane life). Most experimenters tell us they mainly rely on language style and subtle emotional differences to judge whether an agent is human or AI. Still, the overall feeling is already very close to reaching a degree of authenticity.

Furthermore, users interact with embodied agent simulation systems primarily through natural language, which is very friendly to researchers from all fields, especially non-expert users, as they can easily observe and study an agent's personality, thoughts, and behaviors directly. We believe the outputs from each module in this process can greatly aid research in design studies, psychology, sociology, anthropology, and other disciplines.

Perhaps in the future, developed AI can construct highly abstract, high-dimensional models of human personality development. AI can predict a person's personality development changes more accurately and directly without the need for us to build a simulation system to continuously supply the environment and events needed for this prediction process. The "one-click prediction" technology model may become the main technical route for future research on individual development and even most human behavioral psychology.

However, we believe sandbox-style simulation systems still hold very positive value. Although black-box AI models are used in the system to execute specific units such as characters' thinking and behavior, when we look at the entire system, the journey from planning to behavior, from behavior to thinking, and finally to personality development, is an interpretable and traceable "white-box mode." Compared to letting future-developed AI directly provide answers, we believe the process information and forward-deriving logic provided by the simulation system will better enhance the interpretability and transparency of HAI systems. Users can also better understand the system and the human behavioral psychology explored behind it through interaction with such a system. 
\subsubsection{Limitations of large model foundational capabilities on agents}
During the actual design and testing of Evolving Agents, we also discovered several limitations in applying embodied agents. First, due to the positive bias in the content of large language models, the growth trajectory of agents simulated over time may be overly positive, which could lead to a convergence in the development direction of agents as time progresses. Secondly, the state of Agents simulated by large language models is akin to role-playing, which may prevent Agents from fully understanding the complex and subtle emotions humans experience in response to real-life events, often resulting in emotions that are too positive or somewhat confusing. Finally, since real humans are more contradictory, complex, perplexed, and repetitive, whether it is possible to simulate humans in a more complex and realistic manner may require further in-depth research by psychologists and computer scientists.
\subsection{Ethics and Societal Impact}
As the technology of large language models and agent architectures evolves, this field will also trigger more reflection on ethics and morality. Firstly, designers and researchers from other fields must be cautious of becoming overly reliant on embodied agents during the usage process. These agents can only simulate humans' thinking and behavior processes but cannot fully display the complexity and diversity of real humans. Due to the limitations of the architecture, the characteristics of underlying models, and the differences between human cognition and AI cognition, AI will inevitably have instances of inaccurately simulating real humans. Therefore, we suggest using embodied agents only as divergent tools in the early stages of design or research. In actual design and in-depth study, real humans should still be the subject of research to truly understand people.

Secondly, for the whole society, the increasing capabilities and human-likeness of embodied agents will make it increasingly difficult for humans to distinguish between AI and people. Humans may form broader and more profound emotional connections with AI agents, leading to changes in social relationships. This may inspire us to reflect on the connections between humans and society. Although we will inevitably enter a world where virtual and reality merge, we still recommend employing more means and methods to help humans focus more on real life and establish more connections with people around them.
\section{CONCLUSION}
This paper introduces Evolving Agents, interactive computational agents that simulate the diverse and dynamic human personality. We describe a novel architecture based on psychological model mapping, which can perform embodied behaviors according to differentiated personalities in an interactive simulation system and undergo continuous personality evolution based on external information during the interaction process. Evaluations in the sandbox simulation system show that our architecture creates agents with believable and perceptible personalized behaviors and a bidirectional evolution process of personality behaviors. Additionally, through workshops, we have preliminarily verified that such systems have a significant design inspiration during interaction with designers, demonstrating the development potential of embodied agents. Looking ahead, we suggest that Evolving Agents can be applied in the early stages of design and related research in psychology, sociology, and anthropology.
\section{ACKNOWLEDGEMENTS}
This article was funded by the Important International Academic Conference Award Fund for Graduate Students of Tongji University and the Key Technology Research and Demonstration Project of Hydrogen Town Fuel Cell Integrated Energy System\cite{allportPatternGrowthPersonality1961}.

%%
%% The next two lines define the bibliography style to be used, and
%% the bibliography file.
\bibliographystyle{unsrt}
% \bibliography{UistReference_1}
%\bibliography{test1_2}
\bibliography{test1}
\end{document}